\documentclass{aa}
\usepackage[utf8]{inputenc}
\usepackage{slashbox}
\usepackage{natbib}
\usepackage[colorlinks]{hyperref}
\hypersetup{
    citecolor=[rgb]{0.04, 0.20, 0.58},  
    linkcolor=[rgb]{0.86, 0.078, 0.235},
    urlcolor=[rgb]{0,0,1}
}
\usepackage[]{color}
\definecolor{Red}{rgb}{0.9,0.17,0.31}
\usepackage[normalem]{ulem}
\usepackage{soul}

\definecolor{Green}{rgb}{0.0, 0.5, 0.69}

\title{Probing the azimuthal environment of galaxies around clusters}
\subtitle{from cluster core to cosmic filaments}

\author{
  C.~Gouin\inst{1}\thanks{E-mail:~\tt{celine.gouin@ias.u-psud.fr }},
N. Aghanim\inst{1}, V. Bonjean\inst{1}, M. Douspis\inst{1}
}
\institute{
Université Paris-Saclay, CNRS, Institute of Space Astrophysics, 91405, Orsay, France
\label{inst1}
}
\date{\today}

\abstract
{
Galaxy clusters are connected at their peripheries to the large-scale structures by cosmic filaments that funnel accreting material.
These filamentary structures are studied to investigate both environment-driven galaxy evolution and structure formation and evolution. 
In the present work, we probe in a statistical manner the azimuthal distribution of galaxies around clusters as a function of the cluster-centric distance, cluster richness, and star-forming or passive galaxy activity.

We performed a harmonic decomposition in large photometric galaxy catalogue around $6400$ SDSS clusters with masses $M>10^{14}$ solar masses in the redshift range of $0.1<z<0.3$.
The same analysis was performed on the mock galaxy catalogue from the light cone of a Magneticum hydrodynamical simulation.

We used the multipole analysis to quantify asymmetries in the 2-D galaxy
distribution. In the inner cluster regions at $R<2R_{500}$, we confirm
that the galaxy distribution traces an ellipsoidal shape, which is more
pronounced for richest clusters. In the outskirts of the clusters ($R=[2-8]R_{500}$), filamentary patterns are detected in harmonic space with a mean
angular scale $m_{\rm mean}= 4.2 \pm 0.1$. Massive clusters seem to have a larger number of connected filaments than lower-mass clusters.
We also find that passive galaxies appear to trace the
filamentary structures around clusters better. This is the case even if the contribution of star-forming galaxies tends to increase with the cluster-centric distance, suggesting a gradient of galaxy activity in filaments around clusters.
}


\authorrunning{Gouin et al.}
\titlerunning{}

\begin{document}

\maketitle
\section{Introduction}

        Galaxy clusters represent the most recently formed cosmic structures in the hierarchical ladder of the structure formation model. These most massive systems are  at the nodes of the underlying large-scale cosmic web composed of filaments, walls, and voids \citep{Klypin1983,Bond1996}.
        The complex network of filaments is well drawn by large galaxy surveys, from the Center for Astrophysics Redshift Catalogue \citep[CfA,][]{deLapparent1986}, the Sloan Digital Sky Survey \citep[SDSS,][]{sdss}, the Two-degree-Field Galaxy Redshift Survey \citep[2dF,][]{Cole2005} until recently with the Dark Energy Survey \citep[DES,][]{Abbott2016}.
        In this large-scale picture, galaxy clusters result from merging events \citep{Tormen2004} and continue to grow by accreting galaxies, gas, and small groups. 
    These largest virialised systems represent anchors in the overall large-scale structure, actively studied as cosmological probes, via for example the concentration-mass relation \citep[see e.g][]{Buote2007,Mandelbaum2008,Okabe2010} and cluster counts \citep[see e.g.][]{Planck2016,Salvati2018}. 
    The measurement of such cosmological probes often assumes spherical symmetry to describe matter distribution inside galaxy clusters.
    It is however established from both theory and observations that clusters are better approximated as triaxial objects \citep[for a review on cluster shape see][]{Limousin2013}. 

 The triaxiality of these massive systems is inherent to the gravitational collapse of primordial density fluctuations modelled as a Gaussian random field \citep{Bardeen1986,White1979,Bond1996}.
It has been confirmed by N-body simulations that dark matter haloes  of galaxy clusters are approximately prolate ellipsoids \citep[see for examples,][]{Warren1992,Cole1996, Jing2002,Suto2016,Vega-Ferrero2017}. Recently, the non-sphericities of stellar, gas, and dark matter components in galaxy clusters have been quantified in hydrodynamical simulations \citep{Okabe2018}.
The ellipsoidal shape of clusters has been estimated via observables such as galaxy distribution \citep[e.g.][]{Paz2006, Shin2018}, X-ray surface brightness \citep[e.g.][]{Kawahara2010, Lovisari2017}, gravitational lensing \citep{Evans2009, Oguri2010, Clampitt2016B, vanUitert2017}, and the Sunyaev Zel’dovich (SZ) effect \citep[e.g. ][]{Donahue2016}.
Any departure from spherical matter distribution has a significant impact on the inferred halo masses and mass profiles \citep[see e.g.][]{Corless2009}.
For example, the mass-concentration relation is biased by both the halo triaxiality and the presence of substructures within the host halo virial radius \citep{Giocoli2012_concentration}.
Exploring the non-sphericities of matter distribution inside these large over-dense regions is thus crucial for accurately using clusters as cosmological probes.
In our work, we focus on quantifying the level of anisotropy in galaxy distribution from cluster centres to their external regions.

   Deviations from spherical symmetry are expected to increase at cluster outskirts \citep[as measured by][ for example]{Eckert2012}. Indeed, at large radii, infalling material is accreted by clusters in a non-isotropic way along filamentary structures \citep[see e.g.][]{Ebeling2004}.
   Several megaparsec to the cluster centre, assumptions on dynamical equilibrium do not hold anymore \citep{Diaferio1999}, and we might expect accretion shocks \citep[see e.g.][]{Molnar2009}. We can define cluster outskirts as radial ranges from the observational limit for X-ray temperature measurements ($\sim R_{500}$) up to few viral radii \citep[for a review on cluster outskirts, see][]{Reiprich2013}.
   Structure formation effects should be widespread in these outer cluster regions and are therefore ideal places to refine our understanding on the growth of structures. For example, the number of connected filaments around clusters depends on the growth factor and on the dark energy equation of state \citep{Codis2018}. Dark energy is expected to stretch the cosmic web and to induce a disconnection of cosmic nodes with cosmic time \citep{Pichon2010}. The connectivity of galaxy groups can also be used as a tracer of mass assembly history \citep{Darragh2019}.

    Independent of the topology of the cosmic web, probing galaxy properties around clusters can be used to investigate environmental impact on galaxy evolution.
    A large number of physical effects have been proposed to quench star formation in galaxy clusters such as ram-pressure stripping, starvation, or tidal interactions \citep[for reviews see e.g. ][]{Boselli2006,Haines2007}, whereas physical mechanisms which are acting in the outskirts of galaxy clusters have been not extensively explored.
    Recent observational studies show that the efficiency of galaxies to form stars increases with increasing distances to the filament spines and cluster centres \citep[e.g.][]{Malavasi2017, Chen2017, Kraljic2018, Laigle2018}.
    An important fraction of galaxies appears to be quenched in cluster outskirts, in particular inside the filamentary structures, where the pre-processing of galaxies might take place \citep[see e.g. ][]{Martinez2016,Salerno2019,Sarron2019}. 
    
    In our study, we propose to quantify any departure from spherical symmetry in galaxy distribution from cluster central regions up to connected cosmic filaments (a few virial radii).
    Moreover, by considering separately passive and star-forming (SF) galaxy populations, we investigate the role of cluster environments in shaping galaxy activity.
    The angular symmetries are measured via the multipole decomposition of the 2-D galaxy distribution, as introduced by \cite{Schneider1997}.
    Indeed, the quadrupole moment of weak lensing signal is often used to determine the ellipticity of dark matter haloes \citep[see eg.][]{Adhikari2015,Clampitt2016B,Shin2018}.
    Focussing on the larger scale, \cite{Dietrich2005} used multipole moments of shear lensing map to argue in favour of dark matter filament between clusters A222 and A223. The detection of inter-cluster filaments by stacking quadrupole moments of cluster pairs was also investigated by \cite{Mead2010}. 
    Probing all multipole decomposition seems promising to characterise the averaged filamentary patterns around clusters \citep{gouin2017}.
    Unlike common techniques for comic filament detection, such as by reconstructing the skeleton of filamentary structures \citep[see e.g. DisPerSE,][]{Sousbie2011} or by stacking signals between cluster pairs \citep[see e.g.][]{Clampitt2016A,Epps2017,Tanimura2019}, our method integrates mass distribution at cluster peripheries up to very large scales ($10R_{500}$) without making any assumption about the extension or geometry of connected cosmic filaments.

    The paper is organised as follows.
    Section 2 describes the observational dataset from a large photometric galaxy survey and the mock datasets from a hydrodynamical simulation.
    Section 3 presents the formalism of multipole moments and our method applied on galaxy distribution. In Sect 4, we present the angular features found as depending on the cluster-centric distance, cluster richness, and  (SF or passive) galaxy activity. Finally, we summarise our work and give our conclusions in Sect. 5.

\section{Datasets \label{sect:data}}

In this section we present the different datasets later used to measure galaxy multipole moments around clusters.
The observational dataset, called \textit{case 1}, combines a large photometric galaxy catalogue with an overlapping sample of galaxy clusters.
In addition, we take advantage of a large mock galaxy catalogue from one current hydrodynamical cosmological simulation to control systematics and to deepen the interpretation of galaxy multipole moments around clusters. The full mock galaxy catalogue from the hydrodynamical simulation is called \textit{case 3}, and the reduced mock galaxy catalogue to mimic the observational dataset (\textit{case 1}) is noted \textit{case 2}.
A summary of these datasets is presented Tab. \ref{Tab:case}.

\begin{table*}
\centering
\begin{tabular}{ | c | c c c  | }
\hline
 \textbf{ Name }&  \textbf{ Data } &  \textbf{ Cluster selection }  &  \textbf{ Galaxy selection }   \\
    \hline  \hline
     \textit{Case 1} & WISE $\times$ SCOSMOS  & 6398 WHL clusters  & $ \rm 9.5 < log10(M_{*}/M_{\odot})< 11.5$  \\
 & $ \sim \  28,000 \ \rm{deg^2}$ & richness>20 &   and $0.1<z<0.3$    \\
    \hline 
    \textit{Case 2}  & Magneticum Light-cone &  3216 clusters   &  $ \rm 9.5 < log10(M_{*}/M_{\odot})< 11.5$  \\
     & 1/8th of the sky & $\rm M_{200}>1\times10^{14} M_{\odot}$ & and $0.1<z<0.3$   \\
     \hline
     \textit{Case 3} &  Magneticum Light-cone  & 12779 clusters & no stellar mass cut   \\ 
     & 1/8th of the sky & $\rm M_{200}>1\times10^{14} M_{\odot}$ &$0.04<z<0.45$   \\
       \hline

\end{tabular}
\medskip
\caption{Summary of all the datasets used in this study, the galaxy catalogue and their sky area, the cluster selection, and the galaxy selection.\label{Tab:case}}
\end{table*}

 \subsection{Observational dataset (\textit{case 1}) }

 \subsubsection{WISE $\times$ SCOSMOS galaxy catalogue }
 
The large all-sky WISExSCOSMOS galaxy catalogue \citep{Bilicki2016} is the result of a cross-matching of Wide-Field Infrared Survey Explorer \citep[WISE,][]{WISE} and SuperCOSMOS in optical \citep{SCOSMOS3,SCOSMOS2,SCOSMOS1} sources. 
It contains photometric redshifts for about 20 million galaxies up to $z\sim 0.4$ (with a median redshift $z_{\textrm{med}} \sim 0.2$), which have a normalised scatter of photometric redshifts close to $\sigma_z \sim 0.03$.
From this large photometric galaxy catalogue, we selected galaxies in the redshift range $0.1 \leq z \leq 0.3 $, for which the star formation rate (SFR) and stellar mass ($\mathrm{M}_\star$) are efficiently estimated by \cite{Bonjean2018} with a random forest algorithm. 
The initial mass function (IMF) from \cite{2001Kroupa} is used to compute the stellar mass and SFR in this observational galaxy catalogue.

\begin{figure}
    \centering
    \includegraphics[width=0.42\textwidth]{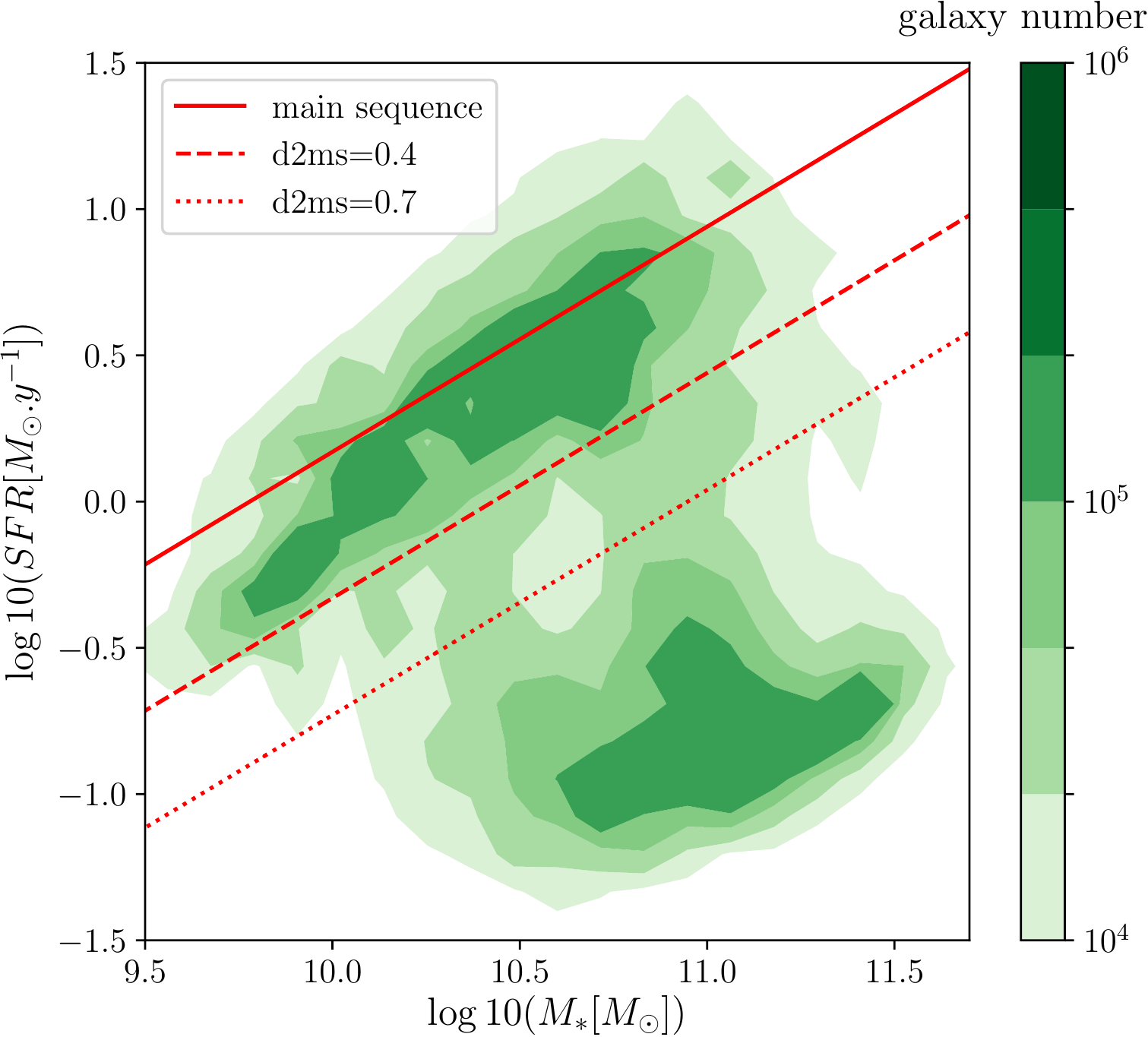}
    \includegraphics[width=0.42\textwidth]{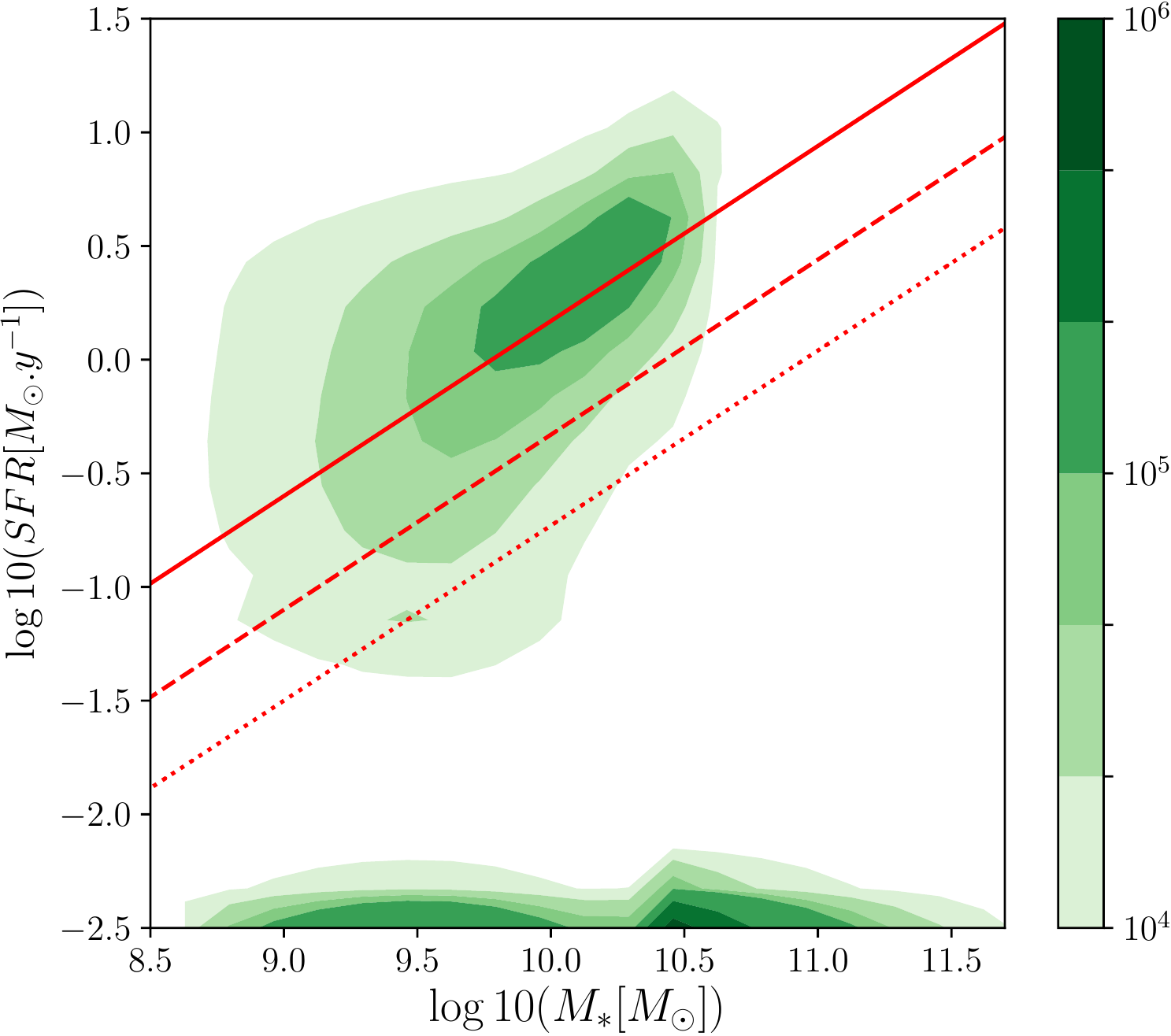}
    \caption{Diagram SFR vs. stellar mass for WISExSCOSMOS datasets (top panel) and for Magneticum galaxies with $0.1<z<0.3$ (bottom panel). The red
solid line is the main sequence of SF galaxies given by \cite{Elbaz2007}, and the dotted red lines show the limit such as $d2ms=0.4$, and $d2ms=0.7$. In the bottom panel, we put artificial mock galaxies with an $SFR=0$  at $\log(SFR)=-2.5$. The zero-SFR galaxies are the result of computational limit in the simulation. \label{fig:sfr_mstar} }
\end{figure}

These two galaxy properties are shown in the top panel of Fig. \ref{fig:sfr_mstar}.  To estimate the type of galaxies, we considered the distance to the main sequence ($d2ms$) on the SFR-$\mathrm{M}_\star$ diagram.
The main sequence of SF galaxies is defined as the SFR-$\mathrm{M}_\star$ relation from \cite{Elbaz2007}, as calibrated on SDSS galaxies. 
As presented in \cite{Bonjean2019}, the distance to the main sequence is used to separate populations of SF ($d2ms < 0.4$), transitioning ($0.4 < d2ms < 1.25$), and passive ($d2ms > 1.25$) galaxies. 

In this work, we separate galaxies into two populations: SF galaxies with $d2ms < 0.4$, and quenched (passive) or in-quenching (transitioning) phase galaxies with $d2ms > 0.4$.
 Indeed, we are interested in quantifying the proportion of SF galaxies in filamentary structures at the cluster outskirts. Galaxies that are not located close to the main sequence are supposed to be quiescent (passive) or have started to be quenched (transitioning).
Moreover, by considering the limit at $d2ms = 0.4$, we insure that the two galaxy populations are in similar proportions. In our redshift range, the catalogue contains 7,249,961 SF and 8,515,574 transitioning/passive galaxies.
To avoid any bias relative to a preferential detection of sources at low or high redshifts (i.e. Malmquist bias), we consider the following stellar mass selection: $9.5<log(M_* [M_{\odot}])<11.5$. That way we ensure a catalogue with almost uniform distributions, by removing very small galaxies that may be preferentially blue ones at low redshift and very massive galaxies that may be only passive galaxies preferentially detected owing to their high luminosities.

\subsubsection{Cluster sample}

Our study focusses on the azimuthal decomposition of galaxy distribution around galaxy clusters. 
Therefore, our dataset is composed of both a galaxy catalogue and an associated overlapping cluster sample.
There are different cluster samples that overlap the WISExSCOSMOS galaxy catalogue in the selected redshift range: either clusters detected by galaxy over-density such as the redMaPPer cluster catalogue \citep{Redmapper}, X-ray detected clusters \citep{MCXC}, or clusters detected via the SZ effect \citep{SZ}.
By measuring the galaxy multipole decomposition around these different cluster samples, we found that the azimuthal shape of clusters are on average independent of the cluster selection. 

We chose to use the cluster sample identified in the SDSS DR8 by \cite{Wen2012}, hereafter called after WHL clusters because of the large size of the sample and its purity level. The completeness of WHL catalogue is estimated to be more than 95\% for clusters with masses $M_{200} > 1.0 \times 10^{14} M_{\odot}$ and redshifts $z < 0.42$, with less than 6\% false detections.
We considered in our study only galaxy clusters with a mass $M_{200} > 1.0 \times 10^{14} M_{\odot}$, which corresponds to a richness threshold of $20$ according to the cluster mass-richness relation calibrated by \cite{Wen2012}.
From the cluster radius $R_{200}$, which encloses 200 times the critical density of the universe, we calculated $ R_{500}$, where $ R_{500} \ = \ 0.65 \ R_{200}$,  by assuming an NFW profile \cite{Navarro1997} with a typical cluster concentration at $c_{200} = 4$ for low-$z$ clusters with $M_{200} > 1.0 \times 10^{14} M_{\odot}$ \citep{Ettori2010, Duffy2008}.

Finally only clusters with a redshift of $0.13<z_c<0.27$ are selected, we considered all the galaxy clusters in the WISExSCOSMOS galaxy catalogue ($0.1<z<0.3$) according to a redshift uncertainty of $\sigma_z \sim 0.03$. 
Following this selection in mass and redshift, we identified 6490 WHL clusters in the observed dataset (\textit{case 1}). We suppressed 95 galaxy clusters which fall (totally or partially) in the masked field of view described in the WISExSCOSMOS galaxy catalogue \citep[][]{Bilicki2016}.

\subsection{Mock datasets (case 2 \& 3)}

In order to test systematics and accurately predict multipole moments on galaxies, we also measured these on the mock galaxy catalogue from the light cone of the \textit{Magneticum Pathfinder}\footnote{http://www.magneticum.org} simulation \citep{Hirschmann2014,Dolag2015}. 
This light cone is a $90 \times 90$ degree field of view from $z\sim0.05$ to $z\sim0.45$. It is constructed from a combination of 13 independent slices and contains more than 30 million galaxies. The simulated galaxies in Magneticum are in overall good agreement with observations in regards to their dynamical properties \citep[see e.g.][]{Teklu2015,Schulze2018,vandeSande2019} and stellar mass function \citep{Hirschmann2014}.
From the Magneticum light cone, we considered two different mock datasets: one which aims to reproduce the observational dataset (\textit{case 2}), and a second larger dataset to better sample the galaxy density field (\textit{case 3}). 

\subsubsection{Galaxy catalogues}

 In \textit{case 2}, mock galaxies are selected in redshift ($0.1<z<0.3$) and stellar mass ($9.5< log(M_* /M_{\odot}) < 11.5$) identically to the actual galaxy catalogue in \textit{case 1}.
 In the second mock dataset (\textit{case 3}), all the galaxies in the Magneticum light cone are considered to improve the sampling of the overall density field. This thus permits us to measure multipole moments precisely with a larger galaxy number. 
 
In any case, in the same way as for the observational dataset, mock galaxies are separated between SF and passive/transiting galaxies, by considering the distance to the main sequence in the SFR-$\mathrm{M}_\star$ diagram \citep{Bonjean2019}.
The IMF from \cite{Chabrier2003} was used to compute the stellar mass and SFR of simulated galaxies \citep{Hirschmann2014}.
The bottom panel in Fig. \ref{fig:sfr_mstar} shows the SFR-$\mathrm{M}_\star$ diagram for mock galaxies in the redshift range $0.1<z<0.3$. 
There are mock galaxies with $SFR=0$ which are artificially placed at $\log(SFR)=-2.5$ on the SFR-$\mathrm{M}_\star$ diagram. 
These galaxies are the result of a simulation and model resolution which does not allow us to resolve small specific SFRs, similar to galaxies in the EAGLE simulation \citep[see][section 5.2]{Guo2016}.
For these zero-SFR galaxies, the distance to the main-sequence $d2ms$ is extremely large. But given that our threshold to separate SF and passive/transitioning galaxies is at $d2ms=0.4$, we suppose that this does not bias our classification into two galaxy populations. This assumption is tested further in \ref{subsect:SF}.

\subsubsection{Cluster samples}

Galaxy clusters in the light cone have been identified and a large number of their properties have been estimated, such as the cluster redshift $z_{c}$,  mass $M_{500}$ (or $M_{200}$), and size $R_{500}$ (or $R_{200}$).
The \textit{Magneticum Pathfinder} simulation was used to predict different galaxy cluster properties and it was found to match well with observations: the intra-cluster light  \citep{Remus2017}, intra-cluster medium \citep{Dolag2017} and galaxy cluster mass function \citep{Bocquet2016}.

We selected the clusters identically as for observations with a minimum cluster mass $M_{200} = 1.0 \times 10^{14} M_{\odot}$. 
In \textit{case 2}, only clusters in the range $0.13<z<0.27$ are selected to reproduce observations, whereas in \textit{case 3} all clusters in the Magneticum light cone are conserved from $z \sim 0.04$ to $z \sim 0.45$. The large size of mock cluster sample in \textit{case 3}  permits us to accurately measure the statistic of multipole moments around galaxy clusters.

Finally, because consecutive light-cone slices are independent, mock clusters close to the edges of light-cone slices (<$R_{500}$) are removed from the cluster sample. 
In addition, clusters near the transverse limit of the light cone (such as $\theta_{x,y} \pm 2\theta_{500} [deg]<0$ or $\theta_{x,y} \pm 2\theta_{500} [deg]>90$) are also discarded. 

\subsection{Comparison between mock and observed galaxy populations (in \textit{case 1 \& 2})  \label{subsect:SF}} 
    \begin{figure}
        \centering
        \includegraphics[width=0.48\textwidth]{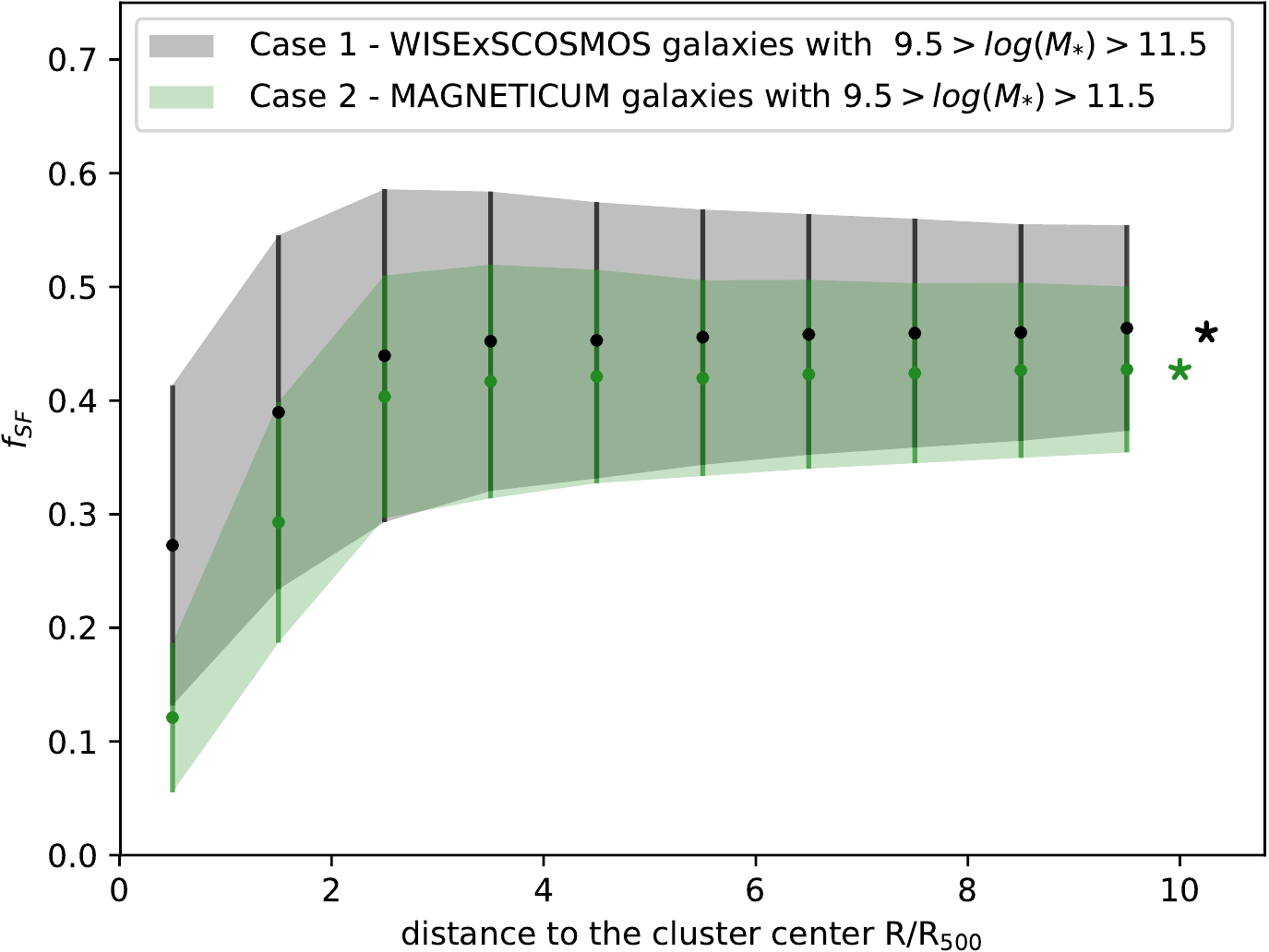}
        \caption{\label{fig:fract_SF} Fraction of SF galaxies as a function of the cluster-centric distance ($R/R_{500}$). The error is the standard deviation of the fraction distribution. The star symbols indicate the mean $f_{SF}$ of the background galaxy field for redshift between $0.1<z<0.3$ and stellar mass selection $9.5<\log(M_{*})<11.5$. }
    \end{figure}
    
   In order to check if our mock catalogue in \textit{case 2} is representative of the observations (\textit{case 1}), we compared their fraction of SF galaxies around galaxy clusters. Indeed, as postulated by \cite{Pintos2019}, the SF fraction is function of the stellar mass, redshift, and environment. Therefore, it is a good estimator to control the selection effects in our two galaxy catalogues.
   The fraction of SF galaxies $f_{SF}$  is calculated as the ratio of the number of SF galaxies to the total number of galaxy inside the aperture $\Delta R$, i.e.
    \begin{equation}
    f_{SF} (\Delta R) = \frac{N_{SF} (\Delta R)}{N (\Delta R)} \,.
    \end{equation}
    In order to probe the fraction of SF galaxies as a function of the cluster distance, the aperture $\Delta R$ is centred on clusters such as $\Delta R  (R_{500})=[0-1],[1-2],[2-3],...$.
    For each cluster, we considered only galaxies in a redshift slice $\Delta z = 2 \sigma_z = 0.06$ and centred on the cluster redshift. The width of redshift slice is discussed further in Sect \ref{sect:multi}. 
    
    Figure \ref{fig:fract_SF} shows the measurement of SF fraction as a function of the cluster distance for observed and mock datasets (\textit{case 1} and \textit{2}). 
    In both cases, the percentage of SF galaxies increases with increasing cluster-centric distance and converges at distance $\sim 3$ to $4 R_{500}$.
    This agrees with the general agreement that galaxy properties converge to those on the field up to $\sim 2-3 $ virial radii  \citep{Ellingson2001, Rines2005,Verdugo2008,Pintos2019}. We note that we define  the SF fraction in the "{background}" as the ratio of the number of SF galaxies to the total number of galaxies in the full galaxy catalogue; redshift and stellar mass selection were discussed before ($0.1<z<0.3$ and $9.5<\log(M_{*})<11.5$).
     The background SF fraction in simulation and observations are in good agreement, showing that our stellar mass selection provides similar galaxy populations.

\section{Aperture multipole moments \label{sect:multi}}

 In this section we present the formalism of multipole moments and the method applied on the different datasets presented in Sect. \ref{sect:data}.
 Aperture multipole moments are used to characterise statistically the anistropies in a 2-D galaxy distribution around galaxy clusters.
 
\subsection{Aperture multipole moment $Q_m$ formalism}
 
\begin{figure*}
    \centering
    \includegraphics[width=0.85\textwidth]{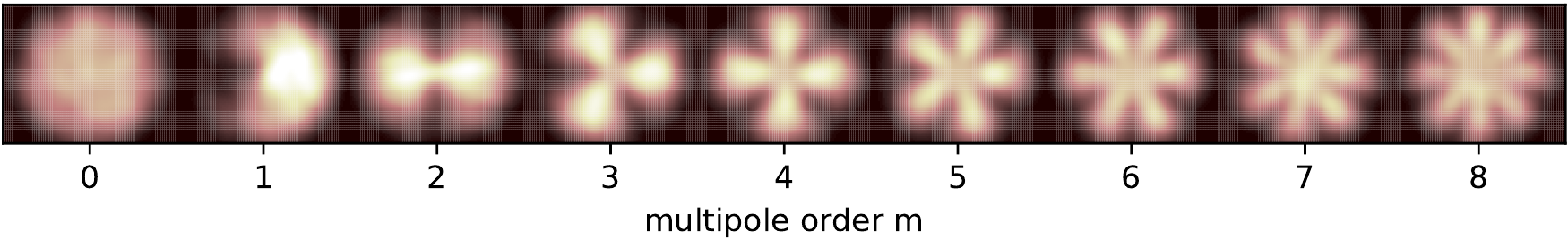}
    \caption{ \label{fig:multi} Illustration of the different 2-D angular symmetries as a function of multipole orders $m$. }
\end{figure*}

\begin{figure}
    \centering
    \includegraphics[width=0.35\textwidth]{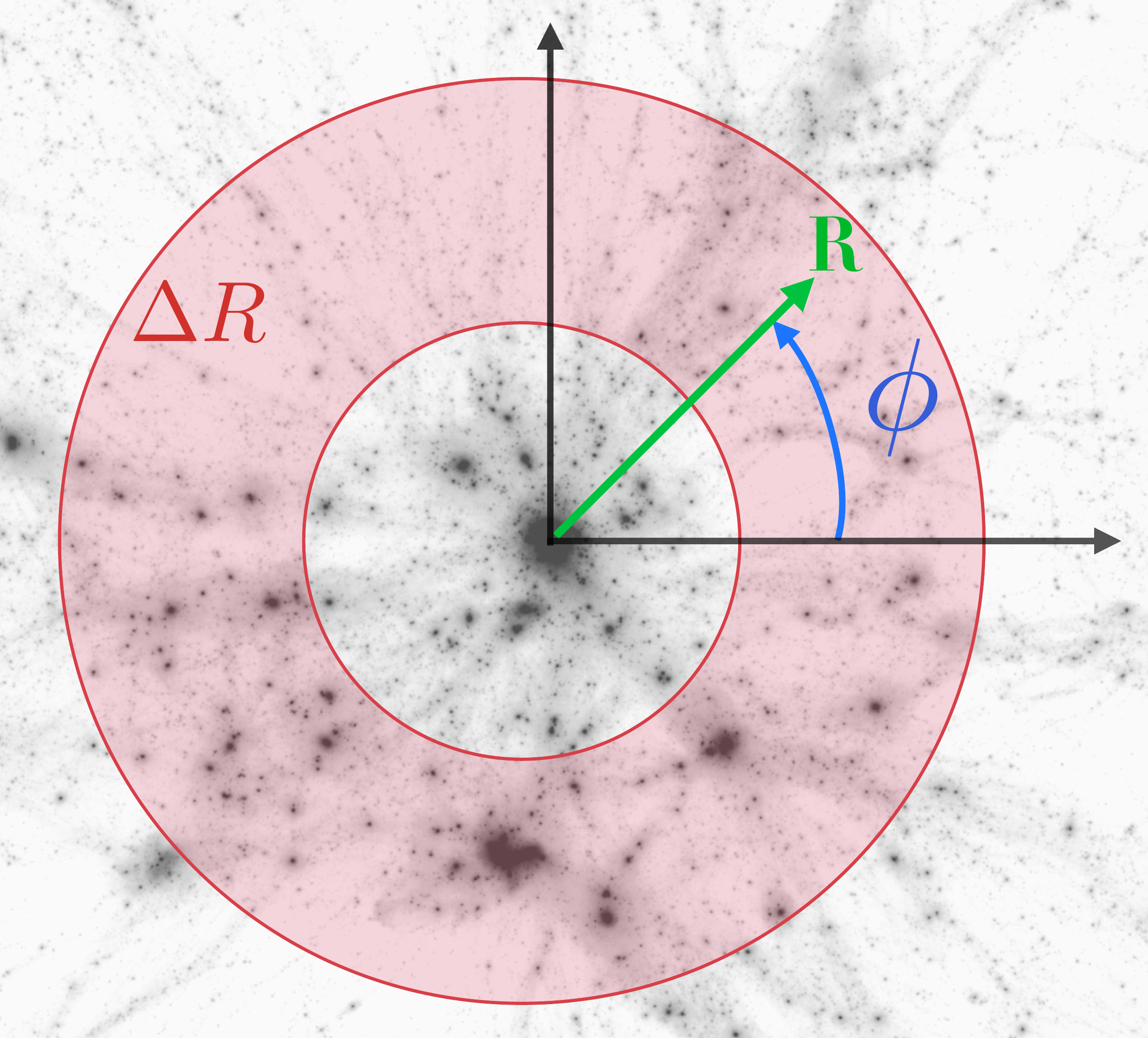}
    \caption{\label{fig:coord} Illustration of the polar coordinate system ($R,\theta$) and a example of radial aperture $\Delta R$ for projected matter distribution centred on a mock galaxy cluster from the Illustris simulation \citep{illustris2014}.}
\end{figure}

 As discussed in the Introduction, the evidence of asphericity of galaxy clusters from simulations and observations has been established for more than 30 years \citep[for a review on cluster shape, see][]{Limousin2013}.
In this work, we propose to evaluate the full azimuthal shape of galaxy clusters from their central to their external regions.
We use decomposition in multipole orders $m$ to highlight possible angular symmetries as illustrated in Fig. \ref{fig:multi}. 
Multipole moments of matter surface density $Q_m$ were first introduced by \cite{Schneider1997}, and are written as
\begin{equation}
     Q_m (\Delta R) = \int_{\Delta R} w(R) R \ dR \ \int_0^{2\pi} d \phi \ e^{im\phi} \ \Sigma (R,\phi) \,,
     \label{eq:multi1}
\end{equation}
 where the polar coordinates in the projected plane are $(R ,\phi)$, the radial aperture is noted $ \Delta R$, and $w(R)$ is an optional radial weight function. Multipole moments applied on weak lensing measurements often choose a weight function close to the mass profile of the lens to maximise the signal \citep[as detail by ][]{Schneider1997}. The aperture can be centred on different structures to probe different anisotropic systems.
 For example, by centring the aperture on massive bridges between cluster pairs, the statistics of the quadrupole $Q_2$ have been proposed to probe inter-cluster filaments from a weak lensing signal \citep{Dietrich2005,Mead2010}.
 In the present work, the aperture is centred on galaxy clusters to characterise filamentary structures at cluster outskirts as illustrated in Fig. \ref{fig:coord}.
 This aperture configuration is promising to identify complex filamentary patterns at cluster peripheries, as has been tested in the N-body simulation by \cite{gouin2017}.
 

\subsection{Statistics of $Q_m$ around galaxy clusters}

Since cosmic filaments have low-density contrasts, we have to average the multipole moments over a large number of clusters to highlight the averaged filamentary patterns around galaxy clusters.
We note $Q^{ \rm cluster}_m$ multipole moments with the condition to centre the aperture on galaxy clusters. We assume that galaxy clusters are located at the density peaks of the underlying density field; we simply refer to this as the peak condition.
By averaging directly the multipole moments decomposition $Q^{ \rm cluster}_m$, the phase information vanishes because we do not align and stack galaxy distribution.
Therefore, we focussed on the statistics of the modulus of the multipole decomposition $\vert Q^{\rm cluster}_m\vert^2 $.  
 We also computed the statistics of multipole moments at random positions $\vert Q^{\rm random}_m\vert^2  $ to highlight the filamentary pattern near clusters that are in excess with
respect to the overall large-scale structures.

Theoretically, the statistics of multipole moments with and without the peak condition is expected to be identical for orders $m > 3$, in a Gaussian random field \citep{Codis2017}.
Beyond this Gaussian picture, high dense regions are supposed to evolve locally more rapidly than the overall cosmic structures in the non-linear regime. Density fluctuations within original shells around cosmic nodes are spherically contracted with the cosmic time due to cluster tidal fields. As a result, harmonic power computed near density peaks is amplified at all angular scales $m$.
We can thus define the harmonic power excess as a normalised statistical estimator, which is the ratio of the harmonic power spectrum centred on galaxy clusters relative to the background,
\begin{equation}
    {\widetilde{Q}}_m = \frac{\langle \vert Q^{\rm cluster}_m \vert^2 \rangle }{\alpha \langle \vert Q^{\rm random}_m \vert^2 \rangle } \,.
    \label{eq:qtild}
\end{equation}
The normalisation factor $\alpha$ represents the boost of amplitude at all modes $m$ around clusters induced by the non-linear matter clustering. Further details are provided in the theoretical demonstration of the boost factor $\alpha$ (with the Zeldovich approximation approach and spherical collapse), and its measurements in an N-body simulation; see Sect 2.3 in \cite{gouin2017}.

\subsection{Measuring of harmonic excess over 2-D galaxy distribution}

For the first time, we propose to measure multipole moments of the 2-D projected galaxy distribution $\Sigma_{gal}$. Previous studies of multipole moments have used weak lensing signal, to probe asymmetry in the total projected mass of gravitational lenses $\Sigma_{tot}$ \citep{Schneider1997,Dietrich2005,Mead2010}.
Using weak lensing signal has the advantage to probe the dark matter potential directly, but it is a low significance signal because it is affected by the intrinsic ellipticity of background sources, and by all the matter content along the line of sight (from the lens to the observer).
In the present study, multipole moments are calculated on galaxy distribution inside redshift slices centred on the cluster redshift $z_c$.  

\subsubsection{Redshift slices}

 Galaxies are attributed to clusters following redshift slices; for example for each WHL (or mock) galaxy clusters, only galaxies within a redshift slice centred on the cluster redshift $z_c$ are used to compute multipole moments.
 As detailed in \cite{Laigle2018}, there is no optimal choice for slice thickness to characterise cosmic filaments in 2-D, but it should be in practice calibrated on the redshift uncertainty. Focussing on the  WISExSCOSMOS galaxy catalogue, the median redshift is $z_{\textrm{med}} = 0.2$ and the scatter is close to $\sigma_z \sim 0.03$. 
 In order to get a constant slice thickness over our small redshift range ($0.1<z<0.3$), we set the slice width equal to twice the typical redshift uncertainty \citep[as in][]{Darragh2019}, such as $\Delta z = 2 \sigma_z \sim 0.06 $.
 For each cluster, galaxies within its own redshift slice, are projected in a 2-D plane and centred on the cluster position. Aperture multipole moments are then calculated on this 2-D galaxy distribution. 

 \subsubsection{Computing galaxy multipole moments}
 Starting with a discrete galaxy distribution projected around a cluster, we can rewrite multipole moments Eq. \ref{eq:multi1} such as
 \begin{equation}  
Q^{\rm cluster}_m   (\Delta R) =   \sum_{j \in z_c\pm \delta z} w(R_j) \ e^{im\phi_j}  \,,
 \end{equation}
 where $z_j$ and ($R_j$,$\phi_j$) are the redshift and polar coordinates of the $j$-th galaxy contained in the redshift slice, respectively, as illustrated in Fig. \ref{fig:coord}. 
 The radial weight function $w(R)$ is defined as a window function which follows the radial aperture $\Delta R$. For each cluster, $\Delta R$ is a function of the cluster radius $R_{500}$. In this way, modulus of multipole moments $\vert Q^{\rm cluster}_m \vert^2 (\Delta R)$ from different clusters with different masses (and radial scales) can be averaged.
 
 To compare harmonic power around clusters with the background galaxy field, multipole moments are also computed around random locations $\vert Q^{\rm random}_m \vert^2$, with the same redshift $z_c$ and aperture $\Delta R(R_{500})$ distribution. This provides random profiles computed on the same sky area and on the same redshift range as cluster profiles on average. Moreover, we consider ten times more random profiles than cluster profiles to reduce the dispersion of $\vert Q^{\rm random}_m \vert^2$. Indeed, after testing different values, we found that ten gives a good balance between a high accuracy on $\vert Q^{\rm random}_m\vert^2$ statistics and a reasonable computational time.

Finally, the harmonic power excess ${\widetilde{Q}_m} $ is calculated by bootstrap re-sampling on the original cluster (and random) multipole moment profiles $\vert Q_m \vert^2$.
 For a set of N cluster (and random) profiles $\vert Q_m \vert^2$, we randomly selected N profiles with replacement and computed the average $ \langle \vert Q_m \vert^2 \rangle $. This bootstrap procedure was iterated 1000 times, and thus provides 1000 re-sampling of $ \langle \vert Q_m \vert^2 \rangle $ for clusters and randoms.
For each bootstrap re-sampling, the normalisation factor $\alpha$ is computed as the ratio of the two asymptotes of the cluster to the random averaged profiles, which are reached around order $m \sim 15$ \citep[as determined in][]{gouin2017}, as follows:
\begin{equation}
     \alpha = \frac{\langle \vert Q^{\rm cluster}_m \vert^2 \rangle_{m>15} }{ \langle \vert Q^{\rm random}_m \vert^2 \rangle_{m>15}} \,.
\end{equation}
 The harmonic power excess ${\widetilde{Q}_m} $ is then directly calculated from the Eq. \ref{eq:qtild}, such as ${\widetilde{Q}_m} $ and its error are derived from 1000 bootstrap re-sampling.

 \subsection{Centring of galaxy clusters}
   The cluster centre has an impact on harmonic power at multipole order $m=1$, in particular inside the virial radius \citep{gouin2017}.
   In our study, we chose to centre apertures on cluster centre, defined as the centre-of-mass in the galaxy distribution. Therefore, we re-calculated the centre by the shrinking circle method for both mock and real cluster samples. Starting from galaxy distribution inside $1.5 R_{500}$, we computed the centre of mass therein, and shrunk the circle by 0.5\%. The centre of the circle is updated at each iteration, until reaching less than four galaxies in the inner circle.
   Applying this centring method, the mean shift of cluster centres is about $0.35 R_{500}$ for observational dataset. It induces a  reduction of the amplitude of $\widetilde{Q^{\textrm{cluster}}_{m=1}}$ (up to 20\% in the central regions). Indeed, the asymmetry characterised by the order $m=1$ reflects simply an over-dense side and an opposite under-dense side in the galaxy distribution (as illustrated in Fig. \ref{fig:multi}).  

  \subsection{Influence of redshift interval on the harmonic power excess}
 For a given cluster, galaxies are either members of the cluster, its environment, or  are from the overall galaxy field. 
 This second galaxy contribution are from the overall large-scale structures in front of and behind the cluster along the line of sight and is smoothed by the photo-z uncertainty.
 Increasing the width of redshift slice tends to attenuate the cluster features, and as a result, it reduces the harmonic power that is in excess to the background galaxy field $\widetilde{Q_m} $. Therefore, in our case we chose the minimal redshift slice thickness, considering the photo-z uncertainty.

\section{Results}

As explain in Sect. \ref{sect:multi}, the normalised galaxy multipole moment spectra allows us to quantify asymmetries in the galaxy distribution around clusters relative to background galaxy distribution.
This statistical estimator is calculated in the three different cases: the full mock galaxy catalogue from the Magneticum light cone (\textit{case 3}), the mock galaxies selected in stellar mass and redshift (\textit{case 2}), and the observed WISExSCOSMOS galaxy catalogue (\textit{case 1}).
In this section, we explore the overall evolution of harmonic power excess $\widetilde{Q_m}$ as a function of the cluster-centric distance, cluster richness, and according to the galaxy population considered (passive and SF).

\subsection{Radial evolution (in \textit{case 3})}

We quantify the level of asymmetry in galaxy distribution as a function of the radial distance to cluster centre $R$.
A visual inspection of N-body simulations and theoretical predictions, as described in Sect. 1, allow us to anticipate the behaviour of matter distribution from cluster cores to their outskirts.
Inside galaxy clusters, matter is supposed to be relaxed and triaxially distributed \citep{Limousin2013}.
From virialised regions of clusters up to their outskirts (typically few $R_{vir}$), matter is anisotropically accreted and funnelled through the cosmic filaments that are connected to clusters \citep{Klypin1983}. 
Finally, far from cluster centres, matter distribution should become statistically identical than the overall large-scale matter density field.

\begin{figure*}[h]
    \centering
    \includegraphics[width=0.96\textwidth]{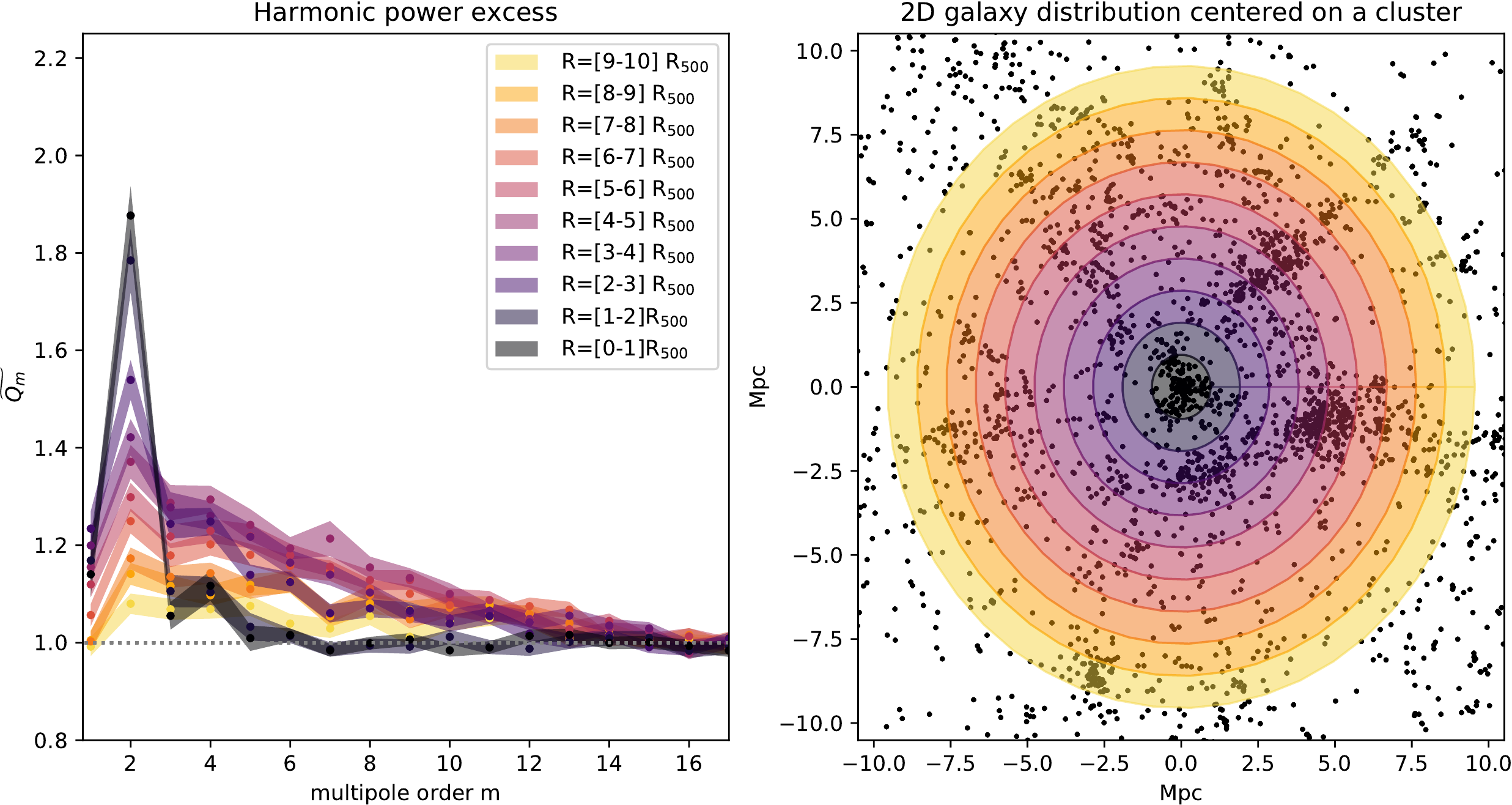}
    \caption{\label{fig:qm_simu} {\it Left panel:}  Harmonic power excess $\widetilde{Q_m}$ for multipole order $m$ from 1 to 17, as a function of different cluster-centric apertures. This is computed by averaging the multipolar moments of the galaxy distribution in \textit{Case 3}.
    Error bars are the error on the mean computed from bootstrap re-sampling.
     {\it Right panel:} Illustration of the different apertures from cluster centre ($[0-1]R_{500}$) to cluster outskirts ($[9-10]R_{500}$). The projected galaxy distribution around one given mock cluster ($ M_{500} \sim 1.3 \times 10^{14} M_{\odot}$ and $z\sim 0.44$) integrated along a redshift slice ($ \Delta z=0.06$) centred on cluster redshift is represented with black points.}
\end{figure*}

Figure \ref{fig:qm_simu} shows the evolution of harmonic power excess $\widetilde{Q_m}$ from the cluster centre to the outskirts in simulations (\textit{case 3}). The different radial apertures around a given mock cluster are drawn on the right panel for illustration. We chose to use the mock dataset in \textit{case 3} to explore this radial evolution of asymmetries because it provides precise results considering the larger number of clusters (increased statistics on multipole moments) and the larger density number of galaxies (reduced noise).

As presented in Fig. \ref{fig:qm_simu}, inside clusters ($R<2 R_{500}$), the harmonic power excess is mainly characterised by the quadrupole $m=2$ (black curves). It means that the 2-D cluster shape is on average elliptical.
We note that there is a small power excess at orders $m=1,3$ and $4$, which could reflect a low level of more complicated asymmetries or substructures.
Beyond $2  R_{500}$, the harmonic power spectrum is distributed on larger multipole orders up to $m\sim12$. The overall level of asymmetries in galaxy distribution increases up to $ \sim 4-5 \ R_{500}$.
These outer cluster regions, above virial radii, are typically the regions in which galaxies infall to galaxy clusters along  the connected filaments \citep[see e.g.][]{Martinez2016}.
At this scale, the large power excess should be therefore the signature of filamentary patterns in harmonic space.
Indeed, \cite{Eckert2012} measure significant deviations from spherical symmetry at cluster outskirts in surface-brightness profile, which suggest accreting materials from the large-scale structures \citep[confirmed by ][]{Eckert2015}.

For apertures distant to the cluster centre such as $ R > 5 \ R_{500}$, the harmonic power excess gradually decreases at all orders $m$ with increasing the cluster-centric distance, down to $\widetilde{Q_m}=1$. As expected, far from the cluster centre, the multipole moments centred on clusters tend to become identical as those computed around random locations.
In other words, the level of asymmetry with and without the condition to be centred on galaxy clusters become identical at very large scales (above few virial radius).

Probing the radial evolution of harmonic power excess in simulation (\textit{case 1}), we conclude that the global ellipsoidal shape of galaxy clusters is approximately contained in $ R < 2R_{500}$, whereas the signature of complex asymmetric structures appears above ($ R > 2R_{500}$), peaks around $4-5 R_{500}$, and becomes negligible at $R \gtrsim 9 R_{500}$.

\begin{figure*}[h]
    \centering
    \includegraphics[width=0.485\textwidth]{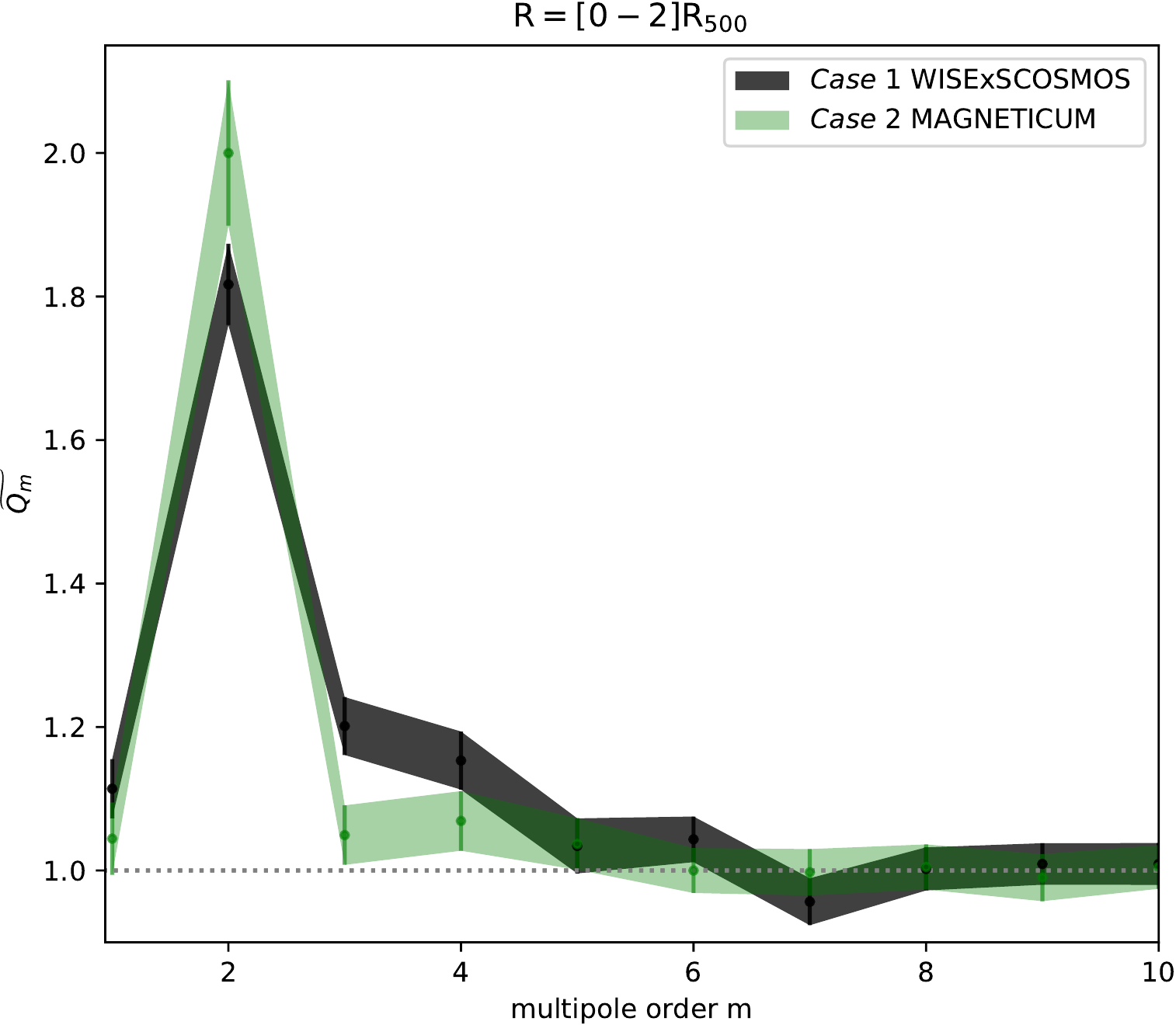}
    \includegraphics[width=0.48\textwidth]{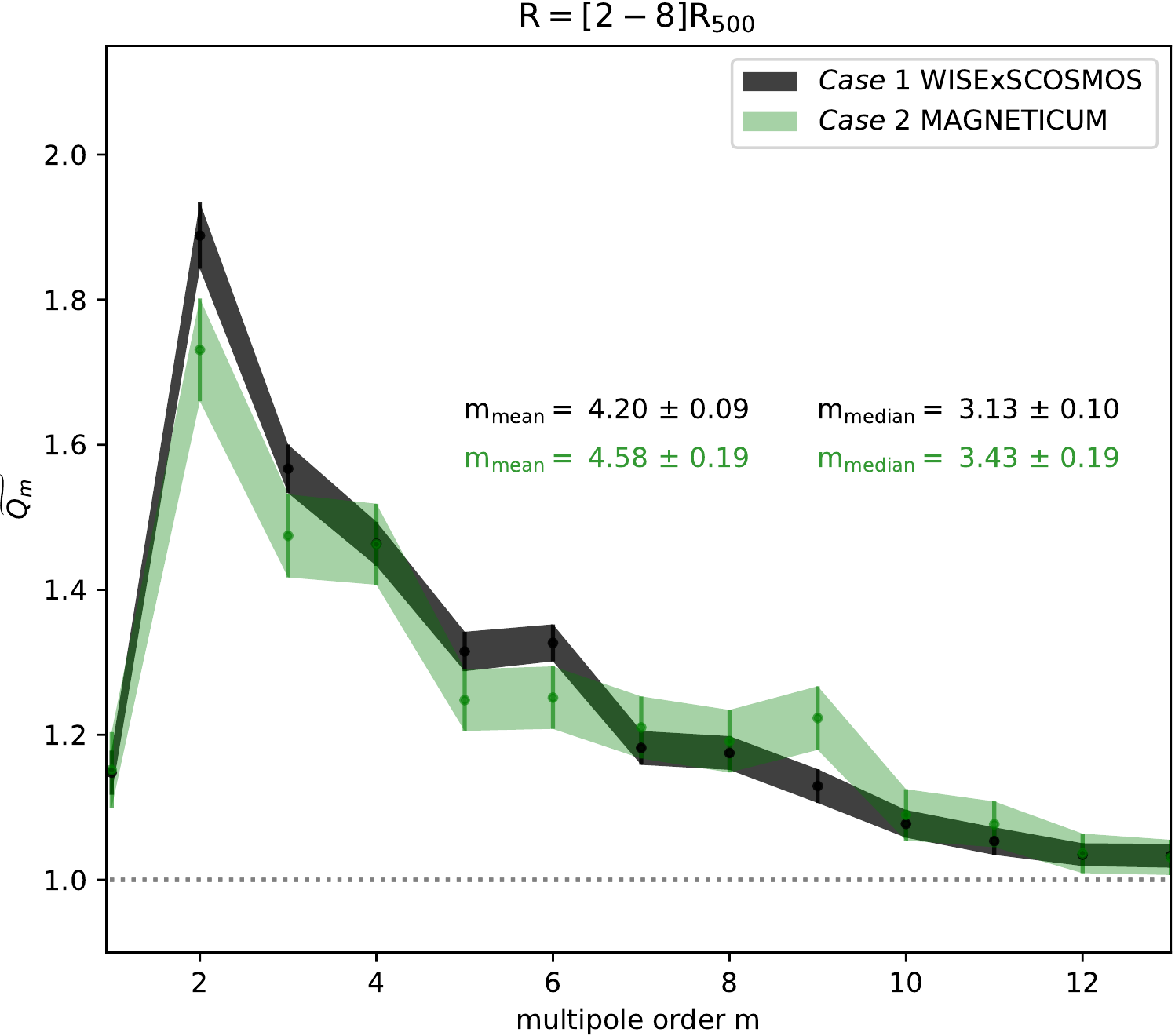}
    \caption{\label{fig:qm_radial_R1R4} {\it Left panel:}   Harmonic power excess $\widetilde{Q_m}$ computed with an aperture $R=[0-2]R_{500}$. 
     {\it Right panel:} Same with an aperture $R=[2-8]R_{500}$. Error bars (and uncertainties on mean and median multipole orders) represent the error on the mean computed from bootstrap re-sampling.} 
\end{figure*}

\subsection{Radial evolution (in \textit{case 1} \& \textit{2})}

In Fig. \ref{fig:qm_radial_R1R4}, the harmonic power excess is measured for two radial apertures, in inner region of galaxy clusters with $R < 2  R_{500}$ (left panel) and in cluster outer regions with $R =[2-8] R_{500}$ (right panel).
The results from the observational dataset in \textit{case 1} are comparable to the harmonic power from mock dataset in \textit{case 2}.
This indicates that our observational results are not affected by systematics and noise contamination.

In the inner region of cluster (left panel), the harmonic power excess is dominated by the quadrupole order ($m=2$).
By probing the full azimuthal shape of observed clusters, we confirm the general agreement that projected galaxy cluster shape is on average elliptical \citep{Limousin2013}. 
In the right panel, the galaxy distribution in the cluster outskirts ($ R_2=[2-8] R_{500}$) presents more than one asymmetry: multipole power excess is spread on different orders from $m=1$ to $m=12$. As discussed previously, this complex harmonic signature must be due to filamentary patterns around clusters. 
Our results obtained from actual galaxy distribution is similar to those found previously from dark matter distribution in N-body simulation by \cite{gouin2017}.
It shows that galaxies are, as expected, good tracers of the underlying matter density field in external regions of galaxy clusters \citep[see e.g.][]{Okabe2008}.

Besides studying the full harmonic decomposition $\widetilde{Q_m}$, we can integrate the harmonic signature over the multipole order $m$ to identify the mean angular scale. 
To do so, we compute a normalised weight applied on $m$ as follows:
\begin{equation}
P_m =\frac{\widetilde{Q_m} -1}{\sum_m (\widetilde{Q_m} -1) } \,,
\end{equation} 
which represents the weight of angular symmetry at each order $m$.
From the distribution of harmonic weight $P_m$, we calculate the median and mean multipole order, noted $m_{\mathrm{median}}$ and $m_{\mathrm{mean}}$, respectively.

 The mean multipole order should depict the mean angular scale in the 2-D galaxy distribution around galaxy clusters. Thus, it should be related with the number of cosmic filaments that are connected to a cluster on average. 
 We find that $m_{\mathrm{median}} \sim 3.13 \pm 0.10$ and $m_{\mathrm{mean}} \sim 4.2 \pm 0.09 $ with the observational galaxy catalogue (\textit{case 1}). 
 In comparison, same analysis from the mock galaxy catalogue (\textit{case 2}) provides similar but slightly higher values of mean (and median) angular scale with $m_{\mathrm{median}} \sim 3.43 \pm 0.19$ and $m_{\mathrm{mean}} \sim 4.58 \pm 0.19 $. 
 In numerical simulation, the number of filaments converging into the node, called cosmic connectivity is around $\sim 3.7$ at $z=0.5$ for 2-D density map \citep{Codis2018}. 
 Observational measurements from \cite{Darragh2019} and \cite{Sarron2019} estimate the mean connectivity around $3-4$ for low-redshift clusters with a mass higher than $M_{200} > 10^{14} M_{\odot}$. These studies measured the connectivity as the number of cosmic filaments that intersect a characteristic radius around clusters (typically $ R_{200}$). In our work, we integrate all the galaxy distribution over a wide radial aperture (from $2$ to $8R_{500}$). This difference in terms of method and integrated aperture can explain the fact that the mean angular scale $m_{\textrm{mean}}$ is slightly higher than the mean connectivity.

\subsection{Correlation between inner and outer cluster regions}

  We aim at investigating the correlation between global shape of galaxy clusters ($R<2 R_{500}$) and the filamentary patterns measured in outer regions ($2R_{500}>R>8R_{500}$).
  Theoretically, the large-scale tidal field and the shape of the density peak are correlated \citep{Bond1996}.
  Previous studies have found a high degree of alignment between the elliptical core of galaxy clusters and their overall environments in N-body simulations \citep[see e.g.][]{Lee2007}  and observations \citep[see e.g.][]{Einasto2018S}. In general, the principal axes of dark matter haloes tend to be aligned with large-scale filaments \citep{Bailin2005, Altay2006, Patiri2006}.
  In particular, \cite{Altay2006} postulated that the alignments of cluster-size haloes are mainly caused by anisotropic merging and infalling of material along filaments. 
    
\begin{figure}[h]
    \centering
    \includegraphics[width=0.42\textwidth]{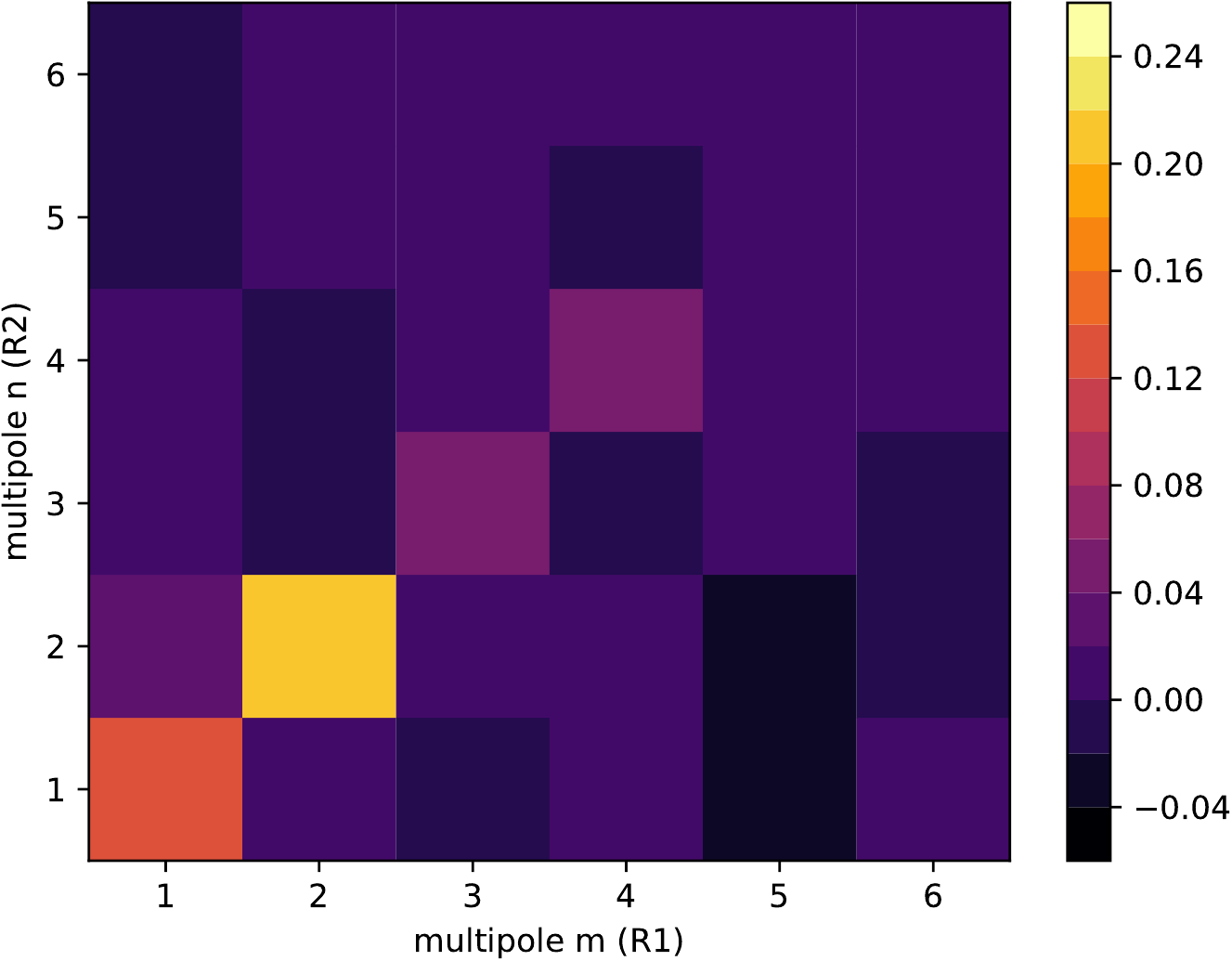}
    \caption{Coefficient of correlation $ \mathcal{C}_{m,n} (R_1,R_2)$ with the two radial apertures $R_1 =[0-2] R_{500}$ and $R_2 =[2-8] R_{500}$. The coefficient at the quadrupole $m=n=2$ mainly prevails. \label{fig:qmqn} }
\end{figure}

Following these works, we can expect that anisotropic directions in the inner and outer cluster regions are related.
To explore this possible alignment between cluster shape and the surrounding galaxy distribution, we correlate multipole moments $Q_m$ with the two different apertures: $ R_1=[0-2]R_{500}$ and $R_2=[2-8]R_{500}$.
 The real part of the correlation coefficients of multipole moments between the two radial apertures $R_1$ and $R_2$ is written as
 \begin{equation}
      \mathcal{C}_{m,n} (R_1,R_2) =  \mathcal{R}e \left( \frac{\langle Q_m (R_1) Q^*_n (R_2) \rangle}{\sigma_{Q_m}(R_2) \sigma_{Q_n} (R_2) } \right) \,.
 \end{equation}
 In Fig. \ref{fig:qmqn}, the correlation between inner ($R_1$) and outer ($R_2$) cluster regions is computed with observed dataset in \textit{case 1}. We find a significant correlation between these two radial apertures at the order $m=n=2$. Similar results are found with the mock galaxy catalogue (\textit{case 2} and \textit{3}).
This high correlation at the quadrupole can be interpreted as a continuity between the ellipsoidal shape of clusters and the filamentary structure at the outskirts of the clusters.

\begin{figure}[h]
    \centering
    \includegraphics[width=0.44\textwidth]{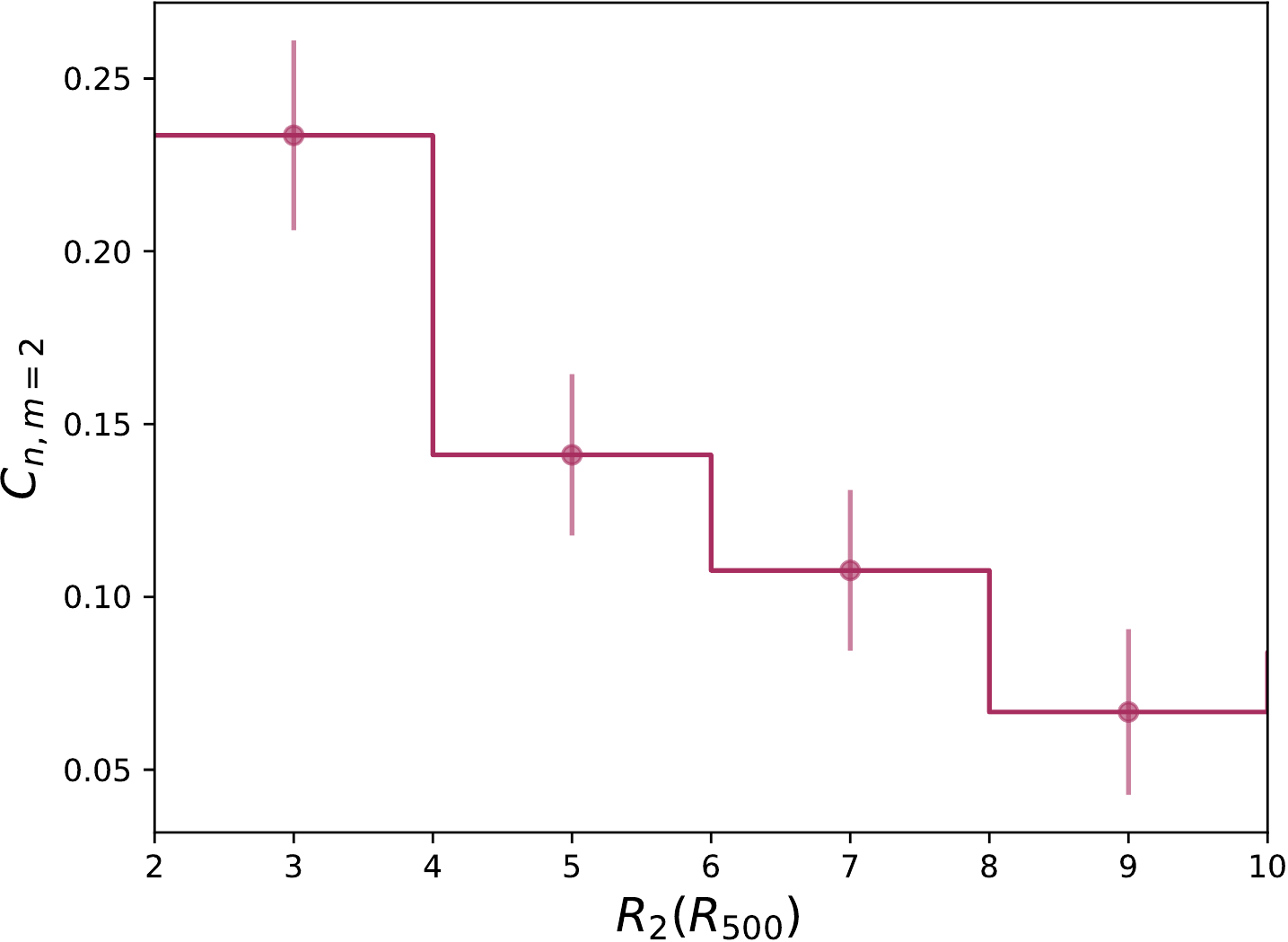}
    \caption{Correlation coefficient $C_{m=2,n=2}(R_1,R_2)$ as depending on the second radial aperture $R_2$, and between fixed at $R_1=[0-2]R_{500}$. \label{fig:Annex_C22}}
\end{figure}

 This agrees with \cite{Codis2018}, who showed that very local density peaks appear just as two ridges, but further away from the centre bifurcation occurs, which increases the number of filaments around the peak.
To confirm this trend, Fig. \ref{fig:Annex_C22} shows the correlation at the orders $m=n=2$ between central region $R_1 = [0-2]R_{500}$, and different annuli distant to cluster centres. As anticipated, the correlation between the cluster ellipsoidal shape and the overall cluster environment decreases with the cluster-centric distance.
 

\subsection{Richness dependence (in \textit{case 1})}

 \begin{figure*}[h!]
    \centering
    \includegraphics[width=0.485\textwidth]{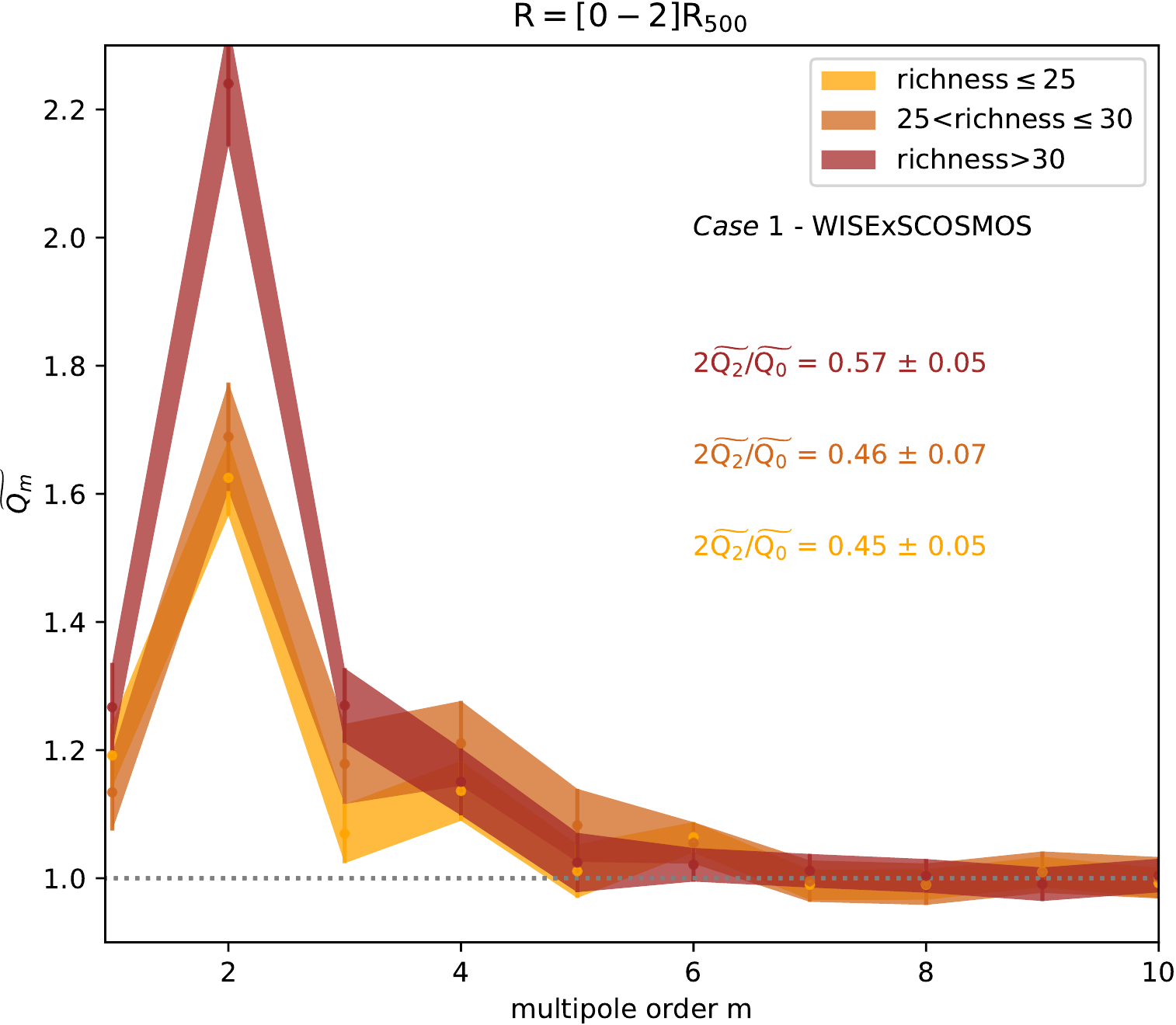}
    \includegraphics[width=0.48\textwidth]{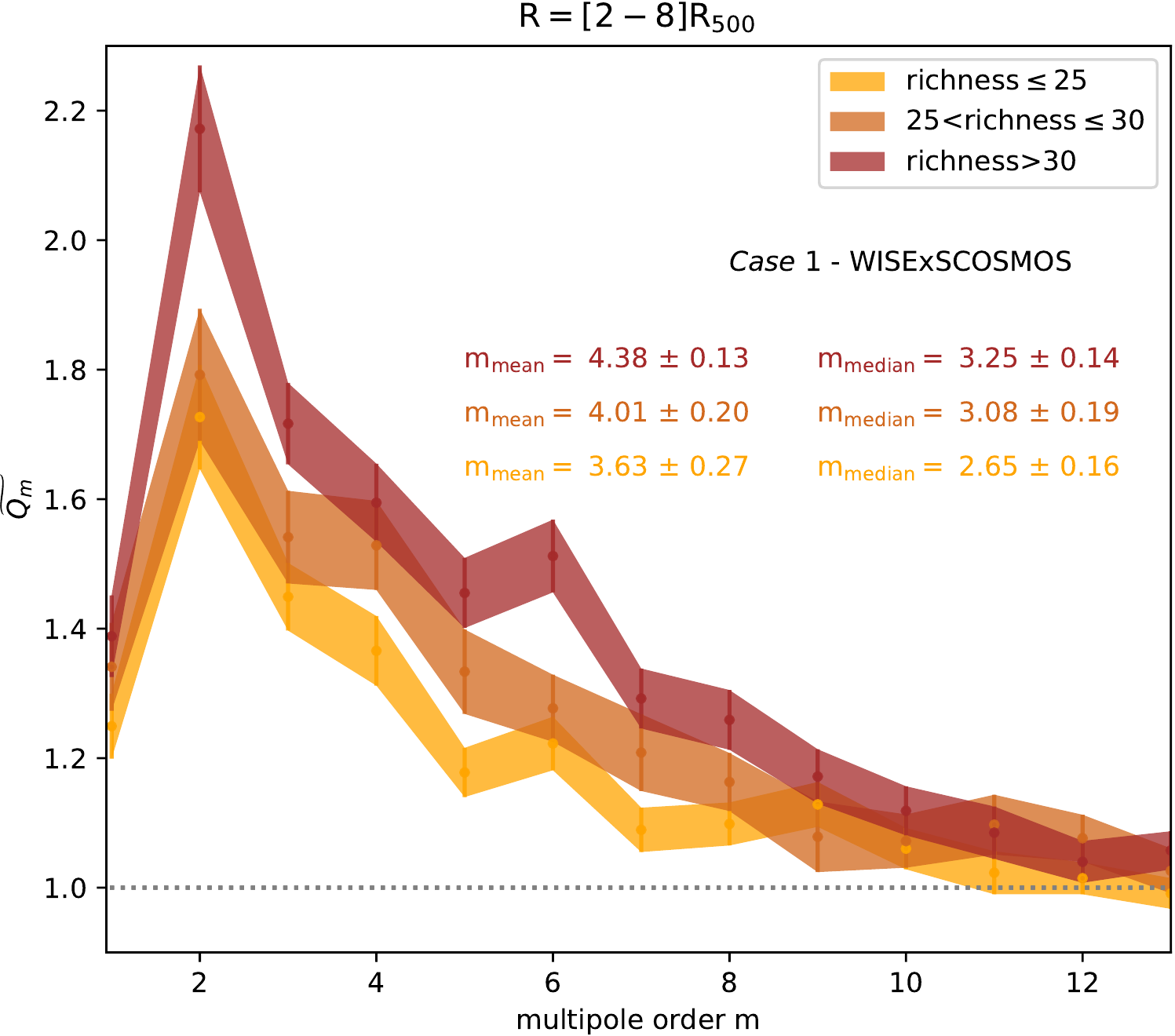}
    \caption{\label{fig:qm_mass_R1R4} Harmonic power excess for the three richness bins and the two radial apertures $R=[0-2]R_{500}$ (right panel) and $R=[2-8]R_{500}$ (left panel). Error bars are computed from re-sampling of the bootstraps.}
\end{figure*}

We investigate the dependence of harmonic power excess with cluster richness on the observational dataset (\textit{case 1}). 
We consider three different bins of cluster richness: $20<richness\leq25$, $25<richness\leq30,$ and $richness > 30$. For each richness bin, the mean cluster mass $M_{200}$ is  $1.25$, $1.58$, and $2.76 \times 10^{14} M_{\odot} $, respectively, following the cluster mass-richness relation from \cite{Wen2012}.  
Fig. \ref{fig:qm_mass_R1R4} shows the harmonic excess for the three cluster richness bins and for the two radial apertures. 

In the inner regions $R<2R_{500}$ (Fig. \ref{fig:qm_mass_R1R4}, left panel), richer clusters ($richness>30$) present a higher harmonic amplitude at the quadrupole ($m=2$) than low-richness clusters. 
This indicates that massive clusters have an elliptical shape that is more marked with a stronger galaxy density contrast on average. To explore the cluster ellipticity, we can use the formalism of \cite{Schneider1991}, which related the harmonic expansion terms of the surface mass density with the ellipticity. For a power-law mass distribution, we can easily show that $Q_2/Q_0  \propto \epsilon /2$. In computing these ratios (quadrupole versus monopole) for the three richness bins, our results suggest that massive clusters have a higher ellipticity than low-mass clusters on average.
 This is in agreement with \cite{Paz2006}, who found that more massive SDSS galaxy groups are consistent with more elongated shapes.
As expected from numerical simulations, the asphericity of dark matter haloes increases with the halo mass \citep[see e.g.][]{Kasun2005,Allgood2006,Despali2014,Vega-Ferrero2017}.

In external regions $2R_{500}<R<8R_{500}$ (Fig. \ref{fig:qm_mass_R1R4}, right panel), we see that the harmonic signature of filamentary patterns is cluster-richness dependent. Rich clusters show a higher harmonic power excess distributed on larger multipole orders $m$ than those with low richness.
Thus, we conclude that the level of asymmetry in galaxy distribution is on average cluster mass dependent. 
The median (and mean) multipole order $m_{\textrm{median}}$ ($m_{\textrm{mean}}$) increases with cluster richness (and hence by mass). 
By assuming that median angular order as a proxy of the connectivity, our results agree with theoretical predictions: massive haloes are expected to be connected to a larger number of filaments than low-mass haloes \citep{Aragon2010,Pichon2010,Codis2018}.  This relation between cluster mass and connectivity has also started to be confirmed in recent observations \citep{Sarron2019,Darragh2019,Malavasi2019}.

In summary, we found that low-richness clusters are more circular, and "less connected" to the cosmic web than richer clusters. In contrast, massive galaxy clusters look more elliptical and present a stronger filamentary pattern at their peripheries. 
These asymmetries in galaxy distribution inside and around clusters might be an indicator of their mass assembly history.
Indeed, \cite{Chen2019} show that a high ellipticity of the intra-cluster medium ($\sim R_{500}$) is connected to a strong mass accretion rate. Focussing on cluster outskirts, \cite{Darragh2019} postulate that a high connectivity in massive groups might be the result of recent merging events.

\subsection{Dependence on galaxy activity (in \textit{case 1} and \textit{2})}

\begin{figure*}[h]
    \centering
    \includegraphics[width=0.485\textwidth]{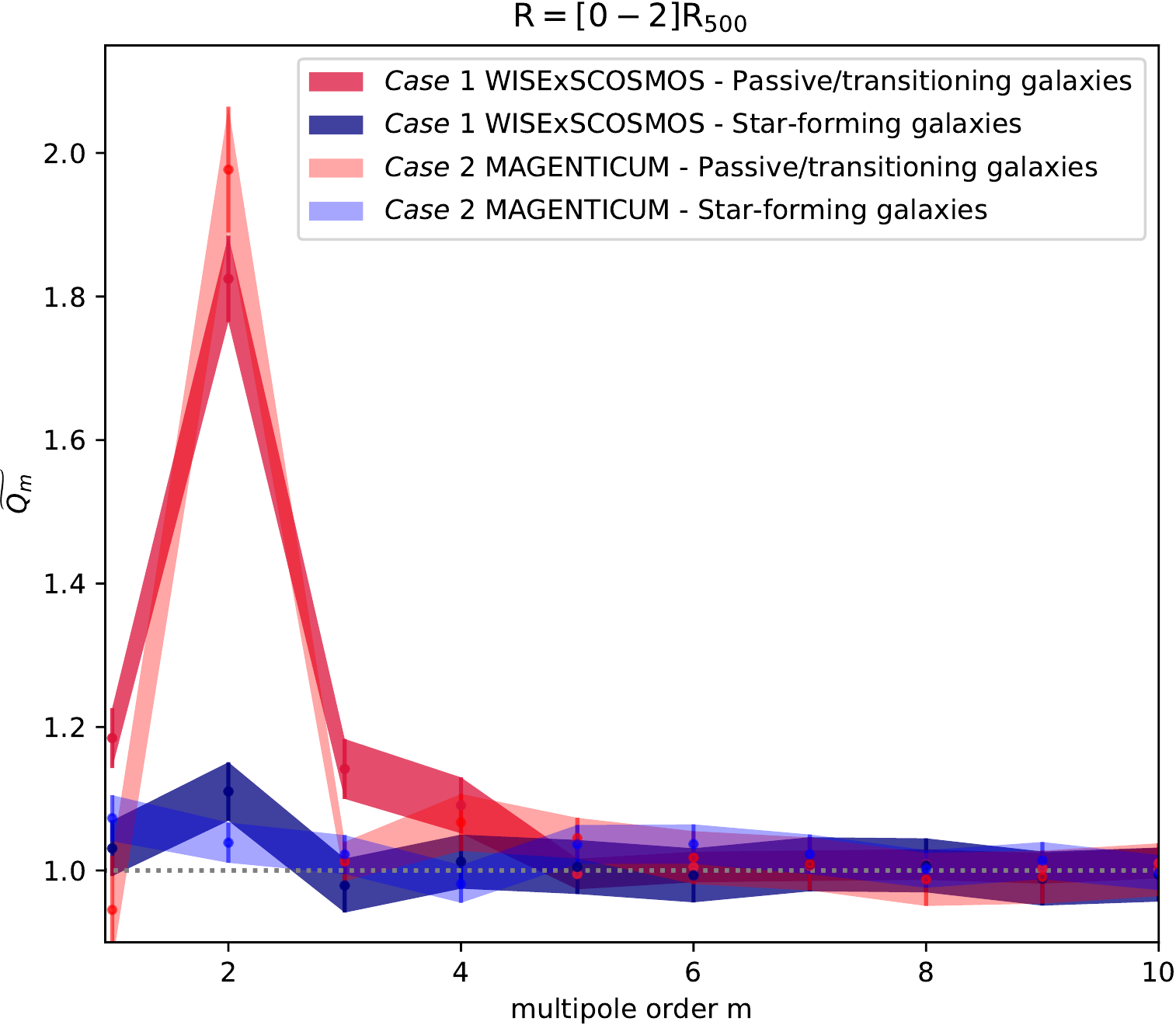}
    \includegraphics[width=0.48\textwidth]{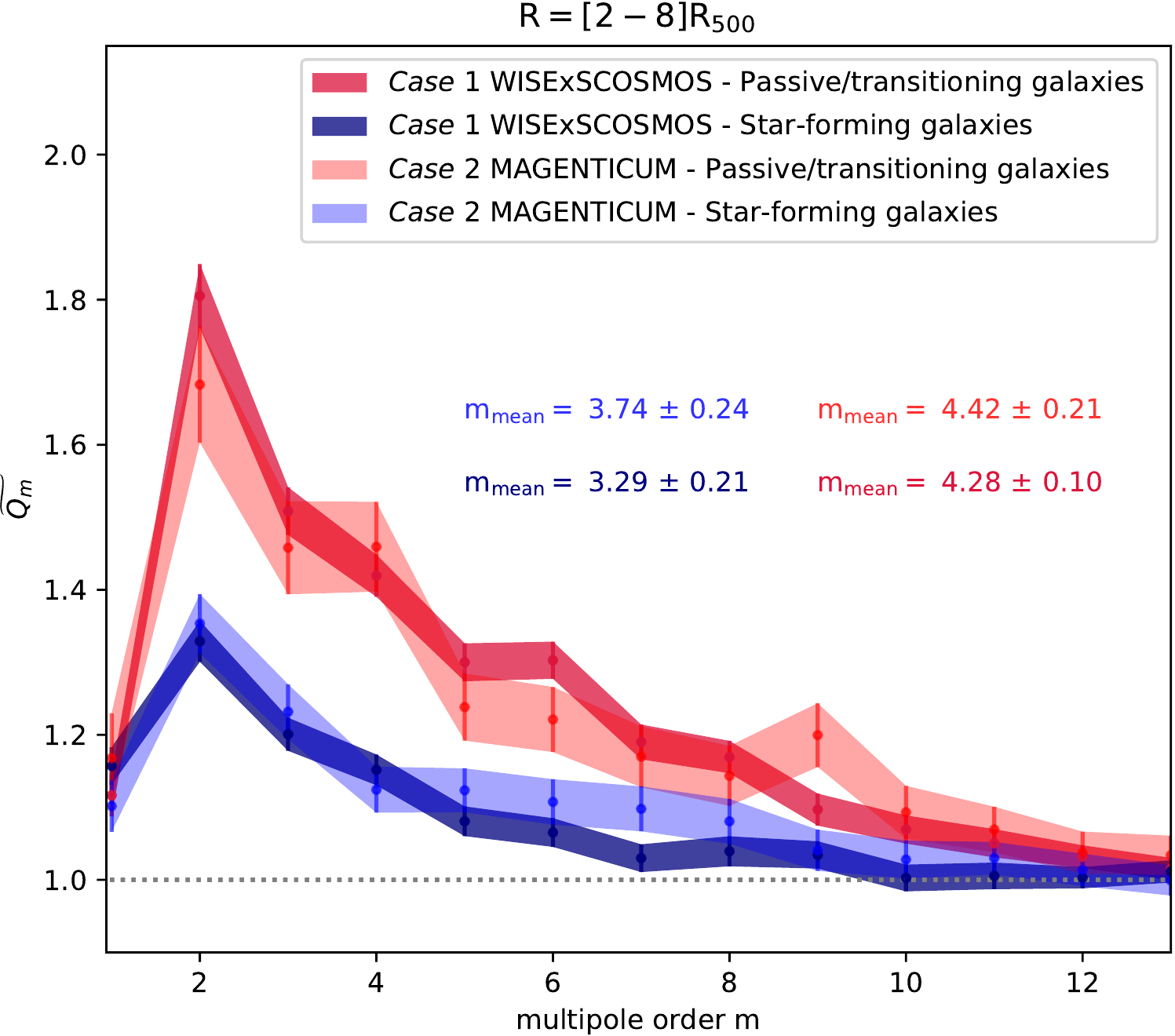}
    \caption{ \label{fig:qm_color_R1R4}  Harmonic power excess computed from two different galaxy population (passive and SF) for the two radial apertures $R=[0-2]R_{500}$ (right panel) and $R=[2-8]R_{500}$ (left panel). Error bars are computed from re-sampling of bootstraps. }
\end{figure*}

\begin{figure*}[h]
    \centering
    \includegraphics[width=0.96\textwidth]{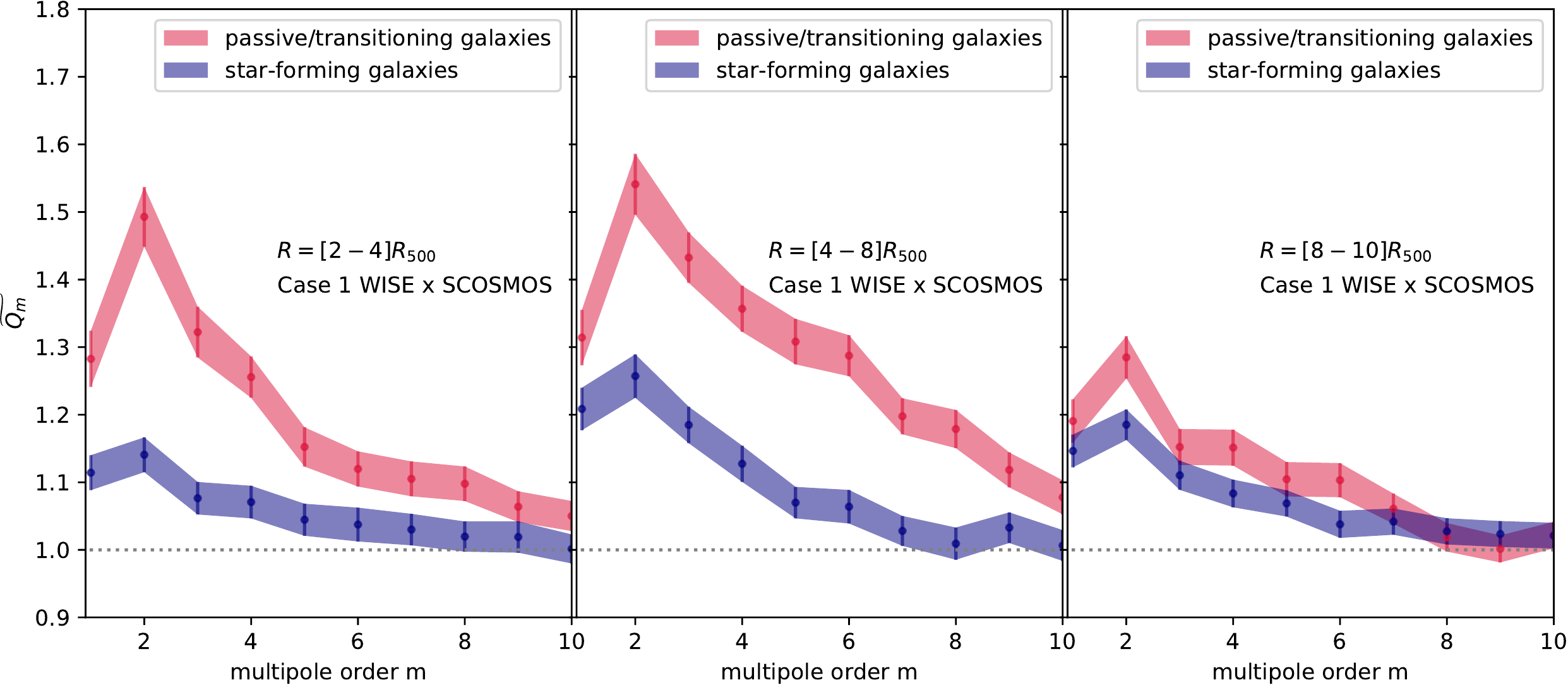}
    \caption{ \label{fig:qm_simu_color}  Harmonic power excess computed from two different galaxy population (passive and SF) for the three radial apertures  $R=[2-4]R_{500}$ (right panel), $R=[4-8]R_{500}$ (middle panel), and $R=[8-10]R_{500}$ (left panel). Error bars are computed from re-sampling of bootstraps. }
\end{figure*}

Passive/transitioning and SF galaxies are treated independently to probe the role of cluster environment on the quenching of star formation in galaxies.
The results are presented in Fig. \ref{fig:qm_color_R1R4}.
Inside galaxy clusters, with $R<2 R_{500}$, the elliptical shape of clusters (characterised by the harmonic excess at $m=2$) is only traced by passive/transitioning galaxies on average in both observational and simulated datasets (\textit{case 1} and \textit{2}).

In cluster external regions ($2>R/R_{500}>8$), harmonic signature of connected filaments is mainly induced by passive/transitioning galaxies with a minor contribution from SF galaxies. This result tends to indicate that filamentary structures around clusters are mainly drawn by passive galaxies, whereas SF galaxies are less concentrated and clustered inside filamentary structures at cluster peripheries. Compared to the result in the region $R<2R_{500}$ suggests that there is a gradient of galaxy activity from the cluster centre to  filamentary structures. 

To confirm this trend, we show in Fig. \ref{fig:qm_simu_color} the radial evolution of harmonic power excess $\widetilde{Q_m}$ as a function of the galaxy types for three radial apertures.
Far from the cluster centres, in the radial annulus $R=[8-10] R_{500}$ (in Fig. \ref{fig:qm_simu_color}, right panel), the harmonic signature is small because the connected filaments become diffuse relative to the overall large-scale structures. Nevertheless, the harmonic power of SF and passive/transitioning galaxies is comparable. 
At the cluster vicinity, for radial aperture $R=[2-4]R_{500}$ (in Fig. \ref{fig:qm_simu_color}, left panel), the contribution from SF galaxies is weak in comparison to the passive/transitioning galaxies.
We conclude that the contribution of SF galaxies in the harmonic signature of connected filaments tends to decrease when approaching the cluster central regions. 
This argues in favour of an increasing gradient of star formation activity inside filamentary structures from cluster centre to the large-scale structure.

Our result is in agreement with \cite{Sarron2019} who show that passive galaxies in cosmic filaments are located closer to clusters than their SF counterpart (for a similar redshift range $0.15<z<0.4$).
Moreover, recent studies indicate that galaxies are systematically more quenched in cosmic filaments around clusters than their counterparts from other isotropic directions \citep[see e.g. ][]{Martinez2016,Einasto2018C,Salerno2019}. In fact, \cite{Lotz2019} find that SF galaxies are in majority quenched during their first orbit around clusters in Magneticum simulation.
Two main possible scenarios can potentially explain a stronger quenching of galaxies in comic filaments at cluster peripheries: either passive galaxies fall more rapidly inside galaxy clusters since they are located closer to filament spine \citep{Laigle2018,Kraljic2018}, or the quenching efficiency is stronger in filaments around clusters as a result of a larger probability for galaxies to merge and to be accreted by small galaxy groups. A pre-processing in galaxy groups falling into massive galaxy clusters might explain the apparent gradient in star formation activity with the cluster distance \citep[see e.g.][]{Bianconi2018}.
Nevertheless, it remains difficult to determine clearly the dominant quenching mechanisms close to high-density environments.

\section{Summary and conclusions}

In this work, we have used the multipole moments of 2-D galaxy distribution to identify angular features around galaxy clusters.
Modulus of multipole decomposition is averaged for a large cluster sample to highlight anisotropies features around clusters statistically.
To quantify the asymmetries around clusters that are in excess to the background galaxy field, we define the harmonic power excess as the ratio of multipole moment spectra centred on clusters to those centred on random locations.
This method permits us to characterise statistically filamentary patterns around clusters in harmonic space.
Unlike the common method for comic filament detection, this method integrates the galaxy distribution inside radial apertures without making any assumption on the geometry or the thickness of cosmic filaments.

This study is performed on the WISExSCOSMOS galaxy catalogue around the $\sim 6400$ WHL galaxy clusters at redshift $0.13<z<0.27$ (and with $richness>20$). Galaxies are selected in redshift slices for each clusters with a constant width $\Delta z = 0.06$ as twice the mean redshift uncertainty.
In addition, same approach has been realised with the mock galaxy catalogue from the Magneticum light cone around mock massive clusters ($M_{200}>10^{14} M_{\odot}$) to control systematics and possible noise contamination.
The harmonic power excess is measured as a function of cluster richness, radial aperture (cluster-centric distance), and (SF and
passive/transitioning) galaxy population.

Mock and real datasets provide similar results, for which the main results are listed above:

\begin{itemize}

\item[(i)]  In cluster inner regions, the projected galaxy distribution appears mainly elliptical on average, i.e. only the quadruple presents a high power excess. 
This confirms triaxial halo shape models and questions the spherical approximation \citep[as expected, see e.g.][]{Limousin2013}. 
Moreover, the quadrupole of galaxy clusters is mass dependent: the elliptical shape of massive clusters is more marked than low-mass clusters on average (based on galaxies enclosed in $R <2R_{500}$). 

\item[(ii)] Considering large radial apertures distant to cluster centres ($R =[2-8]R_{500}$), we detect on average a significant level of anisotropy in galaxy distribution. This large harmonic power excess is supposed to be induced by filamentary patterns around galaxy clusters.
 The mean (median) angular scale is measured around $m_{\textrm{mean}} \sim 4.2 \pm 0.1$ ($m_{\textrm{median}} \sim 3.1 \pm 0.1$).
 These values are in agreement with the number of cosmic filaments departing from galaxy clusters, as measured in observations \citep[see e.g.][]{Darragh2019,Sarron2019}.
 We found that rich clusters have a larger mean angular scale than low-richness clusters, suggesting that they are more connected to the cosmic web. 
 As expected, theoretically, massive haloes  are connected to a large number of filaments \citep{Aragon2010,Pichon2010,Codis2018}.

 \item[(iii)] We probe in detail the evolution of angular features as a function of the cluster-centric distance in simulations. We find that the level of asymmetry in galaxy distribution increases with the cluster distance on average: from an ellipsoidal shape at central regions to complex anisotropic structures at few $Mpc$. Above $\sim 4 R_{500}$, the contrast of asymmetries decreases until  it is identical to overall large galaxy structures.

\item[(iv)] The correlation between the azimuthal galaxy distribution in inner and outer cluster regions is investigated. The elliptical shape of galaxy clusters is significantly correlated with the overall large-scale galaxy distribution outside clusters. This averaged correlation decreases rapidly with the cluster distance. As predicted by \citep{Codis2018}, density peaks appear just as two ridges locally, but further away from the centre, bifurcation occurs and increases the number of filaments. 

\item[(v)] Focussing on the individual contribution of passive and SF galaxies in harmonic power excess, we find that only passive/transitioning galaxies trace the ellipsoidal cluster shape. In cluster outer regions, filamentary patterns are induced mainly by the passive/transitioning galaxies, and by a non-negligible contribution of SF galaxies. This result suggests that SF galaxies are less concentrated in filamentary structures around clusters than passive galaxies, but rather more isotropically distributed.

\item[(vi)] We find that the contribution of SF galaxies in the harmonic filament signature increases when moving away to the clusters. It suggests a gradient in star formation activity inside filamentary structures around clusters. This agrees with recent studies which found that galaxies are systematically more quenched in cosmic filaments around clusters than their counterparts from other isotropic directions \citep{Martinez2016,Salerno2019,Sarron2019}.

\end{itemize}

Future large galaxy surveys such as Euclid and Large Synoptic Survey Telescope (LSST) will allow us to measure galaxy harmonic power around clusters through a large range of redshifts. Unlike this study, which is limited to relatively low-redshift clusters, deep large galaxy surveys will permit us to probe the evolution of harmonic power as a function of cosmic time. As predicted by \cite{gouin2017}, the signature of asymmetries around clusters is expected to increase with the redshift, reflecting the disconnection of galaxy clusters across the cosmic time.

\begin{acknowledgements}
The author thanks the anonymous referee for her/his comments that helped to improve the quality of this paper. The authors thank Alex Saro and Klaus Dolag for making Magneticum galaxy catalogue available. 
The authors also thank all the members of the ByoPiC team (https://byopic.eu/team) for useful comments that have helped to increase the quality of the paper, and Raphael Gavazzi and Clotilde Laigle for fruitful discussions. 
This research has been supported by the funding for the ByoPiC project from the European Research Council (ERC) under the European Union’s Horizon 2020 research and innovation programme grant agreement ERC-2015-AdG 695561. 
\end{acknowledgements}

\bibliographystyle{aa}


\end{document}